\documentclass[a4paper,11pt]{article}

\usepackage{jheppub}

\usepackage{subfig}
\usepackage{xspace}
\usepackage[countmax]{subfloat}
\usepackage{slashed}
\usepackage{longtable}

\usepackage{booktabs}
\usepackage{comment}

\usepackage{bbm}

\allowdisplaybreaks

\def\log{\text{log}}

\def\be{\begin{equation}}
\def\ee{\end{equation}}

\def\zcut{z_{\text{cut}}}

\DeclareRobustCommand{\Sec}[1]{Sec.~\ref{#1}}

\DeclareRobustCommand{\App}[1]{App.~\ref{#1}}

\DeclareRobustCommand{\Fig}[1]{Fig.~\ref{#1}}

\DeclareRobustCommand{\Eq}[1]{Eq.~(\ref{#1})}

\DeclareRobustCommand{\Ref}[1]{Ref.~\cite{#1}}
\DeclareRobustCommand{\Refs}[1]{Refs.~\cite{#1}}

\definecolor{darkblue}{rgb}{0,0,0.5}

\title{Grooming at the Cusp: All-Orders Predictions for the
Transition Region of Jet Groomers}

\author[a,b]{Kees Benkendorfer}
\author[b]{and Andrew J.~Larkoski}

\affiliation[a]{Physics Department, Harvard University, 10 Oxford St, Cambridge, MA 02138, USA}
\affiliation[b]{Physics Department, Reed College, 3203 SE Woodstock Blvd, Portland, OR 97202, USA}

\emailAdd{keesbenkendorfer@g.harvard.edu}
\emailAdd{larkoski@reed.edu}

\abstract{
Jet grooming has emerged as a necessary and vital tool for mitigating contamination radiation in jets.  The additional restrictions on emissions imposed by the groomer can result in non-smooth behavior of resulting fixed-order distributions of observables measured on groomed jets.  As a concrete example, we study the cusp in the hemisphere mass distribution of $e^+e^-\to$ hadrons events groomed with soft drop.  We identify the leading emissions that contribute in the region about the cusp and formulate an all-orders factorization theorem that describes how the cusp is resolved through arbitrary strongly-ordered soft and collinear emissions.  The factorization theorem exhibits numerous novel features such as contributions from collinear modes that can cross hemisphere boundaries as well as requiring explicit subtraction of the limit in which resolved emissions become collinear to the hard core.  We present resummation of the cusp region through next-to-leading logarithmic accuracy and describe how it can be matched with established factorization theorems that describe other groomed phase space regions.
}

\begin{document} 
\maketitle

\section{Introduction}\label{sec:intro}

Analyses on individual jets have matured into a precision theoretical and experimental program to test quantum chromodynamics.  Following the development of the modified mass drop tagger (mMDT) \cite{Dasgupta:2013ihk} and soft drop \cite{Larkoski:2014wba} groomers, theoretical predictions have pushed precision boundaries \cite{Marzani:2017mva,Marzani:2017kqd,Kang:2018jwa,Kang:2018vgn,Caletti:2021oor}, extending resummation of observables on jets to next-to-next-to-leading logarithmic accuracy \cite{Frye:2016aiz,Anderle:2020mxj} and beyond \cite{Kardos:2020gty,Kardos:2020ppl}.  Motivated by these results, unfolded experimental measurements of groomed jet mass have been performed at both the CMS \cite{CMS:2018vzn} and ATLAS \cite{ATLAS:2017zda,ATLAS:2019mgf} experiments at the Large Hadron Collider, and have been directly compared to theory predictions.  These advances set the stage for extraction of fundamental quantities from jet physics, like the strong coupling \cite{Proceedings:2018jsb,Marzani:2019evv,Ringer:2019hdg} or the top quark mass \cite{Hoang:2017kmk}.

All-orders resummation is necessary for precision predictions of observables on highly-boosted jets because a measurement in general introduces a hierarchy between the jet energy and the scale of the measurement; this hierarchy manifests itself as large logarithms in the perturbative prediction of the cross section.  On a groomed jet, the groomer imposes still another scale on the jet, effectively defining the energy of emissions in the jet that survive the groomer and can then contribute to the observable of interest.  Most precision studies thus far have focused on grooming with mMDT and then measuring the jet mass $\rho$ on the groomed jet.  The parameter of mMDT grooming that sets the scale of remaining emissions is typically called $\zcut$ and the highest precision resummation calculations apply in the hierarchical regime in which $\rho\ll\zcut\ll1$.  Phenomenologically and experimentally, this regime is very relevant and can extend over a few decades in the range of observable value, emphasizing the necessity of resummation.  However, with two effective measured scales on the jet, there are other parametric regimes that are relevant for claiming precision over the entire perturbative range of the distribution.

Much less attention has been paid to the regime in which $\rho\sim \zcut \ll1$, which corresponds to the region of the distribution in which the groomer first ``turns on''.  Further, the phase space restrictions imposed on emissions in the jet by the groomer are discontinuous, which results in a cusp or non-smooth point in the distribution when calculated to fixed order in perturbation theory.  Some preliminary analyses have studied this cusp region \cite{Marzani:2017mva,Marzani:2019evv,Larkoski:2020wgx}, but currently, no systematic description exists.  Cusps or transition regions are a general feature of groomers and are present in other algorithms, like pruning \cite{Ellis:2009me} and trimming \cite{Krohn:2009th}; see \Refs{Dasgupta:2013ihk,Dasgupta:2013via}.  In particular, with the assumption that $\rho\sim \zcut \ll 1$, resummation of large logarithms of $\rho$ and/or $\zcut$ would be, in general, necessary to describe this region accurately.  This is related to similar phenomena for other observables that exhibit discontinuous behavior in their distribution away from phase space boundaries at fixed order \cite{Catani:1997xc}.  Our goal in this paper is to provide the systematic, all-orders description of this cusp.

To do this, we first identify how and where this cusp arises at fixed-order on the phase space of the jet.  For simplicity, we focus on hemisphere jets produced in $e^+e^-$ collisions, but discuss how our analysis can be generalized to describe the analogous region for jets produced at a hadron collider.  From this resolved fixed-order description, we can then identify how arbitrary unresolved emissions contribute and present a factorization theorem for the cusp region of the groomed mass distribution within the context of soft-collinear effective theory \cite{Bauer:2000yr,Bauer:2001ct,Bauer:2001yt,Bauer:2002nz}.  This factorization theorem has similarities to the resummation of non-global logarithms and we are able to resum the groomed mass distribution about the cusp to next-to-leading logarithmic accuracy.  Because of the complicated phase space restrictions imposed by the groomer in this regime, extending resummation to higher accuracy may prove prohibitively difficult, but the factorization theorem nevertheless clearly illustrates the physical mechanism responsible for smoothing out the fixed-order cusp.

The outline of this paper is as follows.  In \Sec{sec:obs}, we define the jet grooming algorithm and observables that we measure on our groomed jets.  From these definitions, we can precisely identify the phase space restrictions the groomer enforces in the regime where $\rho\sim\zcut\ll 1$.  In \Sec{sec:fact}, we present the all-orders factorization theorem for resummation about the cusp.  In \Sec{sec:foexp}, we validate the factorization theorem by studying its predictions at leading- and next-to-leading order and comparing to exact or numerical results.  In \Sec{sec:resum}, we present the next-to-leading logarithmic resummed distribution of the cusp region and demonstrate how it can be matched to resummation in other regimes as well as fixed-order.  We conclude in \Sec{sec:concs} and look forward to application of these results to comparison with experimental data.  Calculational details of the anomalous dimensions of the functions appearing in the factorization theorem are presented in the appendices.

\section{Observables and Algorithms}\label{sec:obs}

In this paper, we restrict our focus to $e^+e^-\to$ hadrons collision events.  On these events, we separate the final state particles into two hemispheres, using, for example, the exclusive $k_T$ clustering algorithm \cite{Catani:1993hr} or thrust \cite{Brandt:1964sa,Farhi:1977sg}.  Then, on each of the hemispheres, the soft drop grooming algorithm is applied.  Particles in each hemisphere are reclustered into an angular-ordered branching tree with the Cambridge/Aachen algorithm \cite{Dokshitzer:1997in,Wobisch:1998wt}.  Then, starting with the branches at widest angle, the soft drop requirement on branches $i$ and $j$ is imposed:
\begin{equation}
\frac{\min[E_i,E_j]}{E_i+E_j} > 2^\beta \zcut\sin^\beta\frac{\theta_{ij}}{2}\,,
\end{equation}
where $E_i$ is the energy of branch $i$, $\theta_{ij}$ is the relative angle of branches $i$ and $j$, and $\zcut$ and $\beta$ are parameters of the groomer.  If this inequality fails, then the softer of the two branches is eliminated, and the groomer continues to the next widest-angle branching in the tree that remains.  Once the inequality is satisfied, the recursion terminates and the particles that remain form the groomed jet.  This is referred to as the soft drop groomer for general $\beta$ values \cite{Larkoski:2014wba} and the modified mass drop tagger groomer (mMDT) for $\beta = 0$ \cite{Dasgupta:2013ihk}. 

On the groomed hemispheres, we then measure each hemisphere's scaled mass $\rho$, defined to be
\begin{equation}
\rho \equiv \frac{m^2}{E^2}\,,
\end{equation}
where $m$ and $E$ are that hemisphere's mass and energy, respectively.  Finally, only the larger of the two hemisphere values of $\rho$ contributes to the distribution we predict.  Note that this differs slightly from the usual definition of heavy hemisphere mass because it is normalized by the hemisphere energy and not the center-of-mass collision energy.

\subsection{Schematic Form of Emissions at Cusp}

With this measurement prescription, we note that there are three relevant scales on these groomed jets.  The center-of-mass collision energy $Q$ sets the energy scale of the jets, the grooming parameter $\zcut$ sets the energy scale of the emissions that pass the groomer, and the hemisphere mass $\rho$ sets the relative angle between the emissions that pass the groomer and the hard jet core.  Note that $\zcut$ and $\rho$ are dimensionless, and the corresponding energies that they define are set by their product with $Q$.  So, in identification of different regimes of the groomed jets, we will compare $\rho$, $\zcut$, and 1.  Further, we formally assume that both $\zcut \ll 1$ and $\rho\ll 1$, which implies that the groomer only removes soft emissions from the events and that the masses are dominated by soft and/or collinear emissions.  With this restriction, there are then three possible relationships:
\begin{itemize}

\item $\zcut \ll \rho \ll 1$: In this limit, the emissions that set the hemisphere mass $\rho$ are completely ignorant of the groomer, and the calculation is identical to the standard hemisphere mass.

\item $\rho\ll\zcut \ll 1$: In this limit, all soft, wide-angle emissions are groomed away, and only collinear emissions survive grooming.  The calculation of the hemisphere mass distribution in this limit has been extensively studied \cite{Dasgupta:2013ihk,Larkoski:2014wba,Marzani:2017mva,Kang:2018jwa}.

\item $\rho \sim\zcut \ll 1$: Emissions that contribute to the hemisphere mass may or may not have restrictions imposed on them by the groomer, and this leads to a non-smooth or cusp response of the distribution depending on the relative size of $\rho$ to $\zcut$.  It is this regime that we will focus on in this paper.

\end{itemize}

We can get a broad sense for the structure of events that live in this third regime, where $\rho\sim \zcut\ll 1$, by considering how the groomer and measurement restricts emissions.  Emissions that pass the groomer must therefore have an energy $E$ of the order of $E\sim \zcut Q$, which is a low energy by assumption.  Further, if $\rho\sim \zcut$, then those emissions must be at wide angle to the jet core or near the boundary of the hemisphere; otherwise, the angular factor in the expression for the invariant mass would further suppress its value.  Thus, a hemisphere jet in $e^+e^-$ collisions for which $\rho\sim \zcut\ll 1$ consists of a number of relatively soft, wide angle resolved emissions that pass the groomer, and then arbitrary unresolved soft and collinear emissions off of the resolved emissions.  This configuration of emissions is similar in structure to that which contribute to non-global logarithms (NGLs) for the (more familiar) ungroomed hemisphere mass \cite{Dasgupta:2001sh}.  Here, as in the case of NGLs, the physical configuration of emissions that defines a logarithmic accuracy is difficult to define because the number of resolved soft emissions that can contribute in the cusp region of the distribution is ill-defined, even at leading power in $\rho$ and $\zcut$.  A way forward, however, is to just consider a fixed number of resolved emissions and then sum together all mutually-exclusive phase space configurations to establish the complete result.  General procedures along these lines have been constructed for NGLs \cite{Caron-Huot:2015bja,Larkoski:2015zka,Becher:2015hka,AngelesMartinez:2018cfz}, and we propose a similar line of attack for the groomed hemisphere mass where $\rho\sim\zcut\ll 1$.

\subsection{One-Particle Soft Drop Boundary}

In this paper, we will focus exclusively on those hadronic events in $e^+e^-$ collisions for which there is a single resolved soft, wide-angle gluon that passes the groomer, and then further unresolved, strongly-ordered soft or collinear emissions.\footnote{The analysis presented here for a single resolved gluon that passes the groomer will likely not generalize when more emissions are included.  The non-Abelian structure of the resolved emissions may prohibit factorization of emissions that pass the groomer and strongly-ordered unresolved emissions, without further, explicit measurements; see \Ref{Hoang:2019ceu}.  We thank Aditya Pathak for identifying this subtlety.}  For such events, we can establish the location of the cusp and determine the phase space constraints on the resolved gluon to isolate its contribution around the cusp region.  For a soft gluon with energy $E_s$ and angle $\theta_s$ from the hemisphere jet center, passing the groomer enforces the inequality
\begin{equation}
E_s > E_J \zcut \left(
2\sin \frac{\theta_s}{2}
\right)^\beta\,,
\end{equation}
where $E_J$ is the hemisphere jet energy.  This soft drop constraint is most restrictive for $\theta_s=\pi/2$ when the gluon is right at the hemisphere boundary for which we have
\begin{equation}
E_s > 2^{\beta/2} E_J \zcut \,.
\end{equation}
The contribution to the hemisphere jet mass $\rho$ from this soft emission is
\begin{align}
\rho = \frac{m_J^2}{E_J^2} = \frac{2E_J E_S (1-\cos\theta_s)}{E_J^2} = 2\frac{E_s}{E_J}\,,
\end{align}
for a soft emission right at the hemisphere boundary and assuming that the jet is massless in the absence of the resolved gluon.  The resolved gluon's energy is therefore
\begin{equation}
E_s = \frac{\rho E_J}{2}\,,
\end{equation}
and so grooming begins to affect the hemisphere mass distribution for
\begin{equation}
\rho < 2^{1+\beta/2} \zcut\,.
\end{equation}
The cusp in the groomed hemisphere mass distribution therefore occurs at $\rho = 2^{1+\beta/2} \zcut$, and consideration of more emissions in the hemisphere will smooth it out.

We can write these phase space constraints in a more useful way for calculation and the formulation of the factorization theorem that we present in the following section.  First, we introduce the (dimensionless) light-cone variables $k^+,k^-$ that describe the momentum of the resolved gluon such that
\begin{align}
&k^0 = \frac{Q}{2}(k^++k^-)\,, &k^z = \frac{Q}{2}(k^--k^+)\,,
\end{align}
where $k^0$ and $k^z$ are the gluon's energy and $z$-component of momentum, respectively, and $Q$ is the center-of-mass collision energy.  Then, note that the angle of the gluon to the right hemisphere axis, for example, is
\begin{align}
4\sin^2\frac{\theta_s}{2} = 2 \frac{k^0 - k^z}{k^0}= \frac{4k^+}{k^++ k^-} \,.
\end{align}
Then, the soft drop constraint in these coordinates is
\begin{align}
\frac{Q}{2}(k^++k^-) > \frac{Q}{2}\zcut \left(
\frac{4k^+}{k^++ k^-}
\right)^{\beta / 2}\,,
\end{align}
 or, by rearranging,
 \begin{align}
 k^- > \zcut^{\frac{2}{2+\beta}}(4k^+)^{\frac{\beta}{2+\beta}} - k^+\,.
 \end{align}
 Further, to remain in the right hemisphere we require that
 \begin{align}
 k^- > k^+\,.
\end{align}

These constraints thus far describe any emission that passes the groomer and lies in the right hemisphere.  To restrict to the region around the cusp, where $\rho\sim \zcut$, we need to eliminate the possibility that the resolved gluon has a non-zero contribution when $\rho\ll\zcut$.  This regime is accounted for by the limit in which the gluon becomes collinear to the high-energy quark at the center of the hemisphere jet.   So, to eliminate this regime, we subtract the constraint where $k^-\sim \zcut\gg k^+$ and so the phase space constraints imposed by the groomer to remain in the cusp region are as follows.  For a function $f(k^+,k^-)$ of the lightcone momentum components, the phase space constraints with subtraction are:
\begin{align}\label{eq:sdpsconsts}
\Theta_\text{SD}f(k^+,k^-) &=\Theta\left(k^- - \left( \zcut^{\frac{2}{2+\beta}}(4k^+)^{\frac{\beta}{2+\beta}} - k^+ \right)\right)  \Theta(k^--k^+)f(k^+,k^-)\\
&
\hspace{1cm}-\Theta\left(k^- -\zcut^{\frac{2}{2+\beta}}(4k^+)^{\frac{\beta}{2+\beta}} \right)f(k^+\to 0,k^-)\nonumber\,.
\end{align}
We will mostly be interested in the $\beta = 0$ case for which the groomer reduces to mMDT and the phase space constraint is:
\begin{align}
\Theta_\text{SD,$\beta = 0$} f(k^+,k^-)&=\Theta\left(k^- - \left( \zcut - k^+ \right)\right)  \Theta(k^--k^+)f(k^+,k^-)\\
&
\hspace{1cm}-\Theta\left(k^- -\zcut \right)f(k^+\to 0,k^-)\nonumber\\
&=\Theta\left(k^- - \left( \zcut - k^+ \right)\right)  \Theta(k^--k^+)\left[f(k^+,k^-)-f(k^+\to 0,k^-)\right]\nonumber\\
&\hspace{1cm}+\left[
\Theta\left(k^- - \left( \zcut - k^+ \right)\right)  \Theta(k^--k^+)-\Theta\left(k^- -\zcut \right)
\right]f(k^+\to 0,k^-)\nonumber\\
&=\left[\Theta(\zcut-2k^+)\Theta\left(k^- - \left( \zcut - k^+ \right)\right) \right.\nonumber\\
&\hspace{2cm}\left.+\Theta(2k^+-\zcut) \Theta(k^--k^+)\right]\left[f(k^+,k^-)-f(k^+\to 0,k^-)\right]\nonumber\\
&\hspace{1cm}+\left[
\Theta\left(2k^+ - \zcut\right)\Theta\left(\zcut -k^-\right)\Theta(k^--k^+)\right.\nonumber\\
&\hspace{2cm}+\Theta\left(\zcut-2k^+ \right)\Theta\left(  \zcut -k^- \right)\Theta\left(k^+ +k^- -\zcut \right)\nonumber\\
&\hspace{2cm}\left.-\Theta\left(k^+-\zcut\right)\Theta(k^+-k^-)\Theta\left(k^- -\zcut \right)
\right]f(k^+\to 0,k^-)\,.\nonumber
\end{align}
These $\Theta$-function constraints can be further massaged by noting that in any evaluation of the cross section, we will be integrating over $k^+$ and $k^-$.  With this in mind, consider the first and third constraints on the limiting function $f(k^+\to0,k^-)$:
\begin{align}
&\int dk^-\left[
\Theta\left(2k^+ - \zcut\right)\Theta\left(\zcut -k^-\right)\Theta(k^--k^+)-\Theta\left(k^+-\zcut\right)\Theta(k^+-k^-)\Theta\left(k^- -\zcut \right)
\right]\nonumber\\
&=\Theta\left(2k^+ - \zcut\right)\Theta\left(\zcut-k^+\right)\int_{k^+}^{\zcut}dk^- -\Theta\left(2k^+ - \zcut\right) \Theta\left(k^+-\zcut\right)\int_{\zcut}^{k^+}dk^-\nonumber\\
&=\Theta\left(2k^+ - \zcut\right)\Theta\left(\zcut-k^+\right)\int_{k^+}^{\zcut}dk^- +\Theta\left(2k^+ - \zcut\right) \Theta\left(k^+-\zcut\right)\int^{\zcut}_{k^+}dk^-\nonumber\\
&=\Theta\left(2k^+ - \zcut\right)\int_{k^+}^{\zcut}dk^-\,.
\end{align}
Then, in the final line, there is no assumption made about the relative sizes of $\zcut$ and $k^+$.  Therefore, the subtracted phase space constraints integrated over $k^-$ can be expressed as
\begin{align}\label{eq:psconstsprescrip}
\int dk^-\,\Theta_\text{SD,$\beta = 0$} f(k^+,k^-) &= \Theta(\zcut-2k^+)\int_{\zcut-k^+}^\infty dk^-\,\left[f(k^+,k^-)-f(k^+\to 0,k^-)\right]\\
&\hspace{1cm} +\Theta(2k^+-\zcut)\int_{k^+}^\infty dk^-\,\left[f(k^+,k^-)-f(k^+\to 0,k^-)\right]\nonumber\\
&
\hspace{1cm} + \Theta\left(2k^+ - \zcut\right)\int_{k^+}^{\zcut}dk^-\, f(k^+\to 0,k^-)\nonumber\\
&\hspace{1cm}+\Theta(\zcut-2k^+)\int_{\zcut-k^+}^{\zcut} dk^-\, f(k^+\to 0,k^-)\,.\nonumber
\end{align}

\section{Factorization About the Cusp}\label{sec:fact}

From this leading-order description of a single gluon that probes the cusp of the soft drop groomer, we would like to construct an all-orders factorization theorem that describes arbitrary strongly-ordered soft and/or collinear radiation emitted from this configuration.  This initial configuration consists of a high-energy quark and anti-quark produced from the collision and from which the event hemispheres are defined, and the emitted, resolved, soft gluon which necessarily passes the grooming requirement in its hemisphere.  For concreteness, we assume that the gluon is in the same hemisphere as the quark, but the configuration when the gluon is in the anti-quark hemisphere will have the exact same distribution.  Unresolved collinear emissions can be emitted off of any of the quark, anti-quark or resolved gluon with no grooming requirement enforced because they only lie at small angle.  Unresolved soft emissions can also lie in the angular region between the quark and the resolved gluon and they, too, have no grooming requirement enforced because the resolved gluon at wide angle already passed the groomer.  

As discussed earlier, we make the assumption that there is only one resolved emission off of the final state quark--anti-quark dipole that passes the groomer.  As such, there can be no resolved emissions that pass the groomer in the anti-quark's hemisphere.  Thus, all emissions at wide angles in the anti-quark hemisphere must fail the grooming requirement.  Further, emissions in the quark--gluon hemisphere that lie outside of their angular region must also fail the groomer.  Because we assume that $\zcut\ll 1$, such emissions are necessarily soft, and can be accounted for in the dynamics that are responsible for producing the resolved gluon.  This is similar to construction of the groomed resolved emission function introduced in \Ref{Larkoski:2017cqq}.  With this understanding, any emissions in the anti-quark hemisphere that pass the groomer are necessarily collinear to the anti-quark and therefore result in a hierarchy between its groomed hemisphere mass $\rho$ and the grooming parameter $\zcut$, $\rho\ll\zcut\ll1$.  Such collinear emissions that are sensitive to both the grooming parameter and the hemisphere mass are also low energy, and are described by a collinear-soft function, as introduced in \Ref{Frye:2016aiz}.

As mentioned earlier, this restriction to resummation with a single resolved gluon emission that passes the groomer is similar to techniques developed for resummation of NGLs \cite{Caron-Huot:2015bja,Larkoski:2015zka,Becher:2015hka,AngelesMartinez:2018cfz}.  Specifically, to perform the factorization in this kinematic configuration, in addition to the hemisphere masses, we also explicitly measure the momentum of the resolved gluon.  Soft drop acts non-trivially on the resolved gluon, effectively fixing its energy to be of the order of a fraction $\zcut$ of the center-of-mass collision energy, and its angle from the hard jet cores to be order-1.  Thus, our factorization theorem presented below successfully resums logarithms of the groomed hemisphere jet mass $\rho$, the grooming parameter $\zcut$, and the small energy of the resolved gluon.  In what follows, we will explicitly perform the resummation of these quantities to next-to-leading logarithmic (NLL) accuracy through calculation of the anomalous dimensions of the factorization theorem at one-loop.  However, the final experimental measurement is exclusively the groomed hemisphere mass, so we must integrate over the momentum of the resolved gluon.  This integration then renders the final accuracy of our calculation somewhat ambiguous, because just measuring the groomed hemisphere mass alone does not isolate one resolved gluon.  This is true even at NLL accuracy, which corresponds to terms of the form $\alpha_s^n \log^n \rho$ in the cross section.  However, there is a clear systematic procedure to improve the calculation, and we will present the form of the factorization theorem for two resolved gluons in the conclusions.  Again, this same subtlety is present in calculations of NGLs in the ungroomed hemisphere mass, and there is currently no known method to analytically resum completely at NLL accuracy in one fell swoop with linear renormalization group equations.

\begin{figure}
\begin{center}
\includegraphics[width=7.5cm]{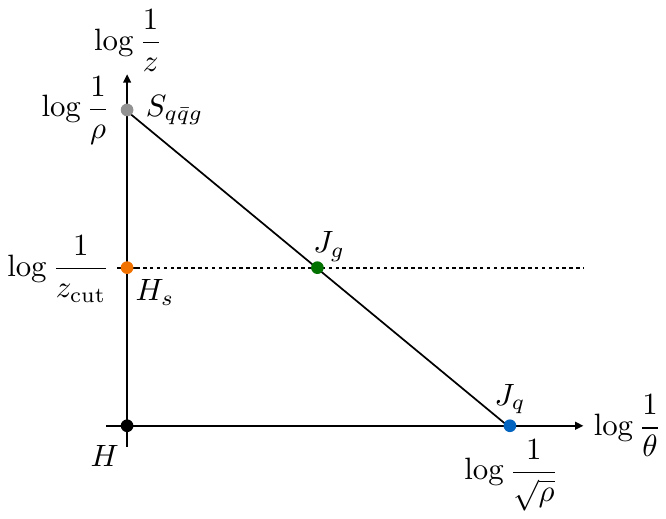}
\hspace{0.5cm}
\includegraphics[width=7.5cm]{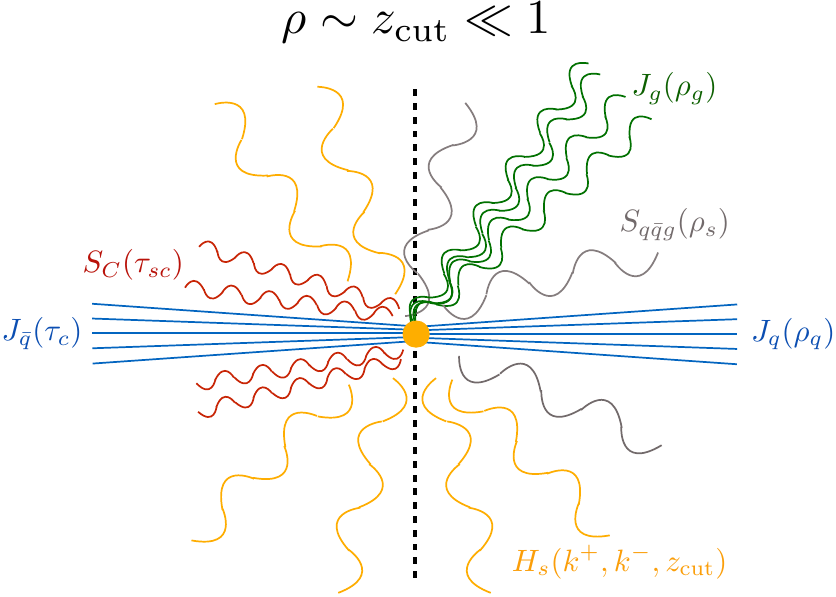}
\caption{
Schematic illustration of factorization theorem for the cusp region of the groomer.  Left: The structure of the factorization theorem for the right hemisphere of the event on the Lund plane, corresponding to the hemisphere with the resolved emission that passes the groomer.  Right: A physical representation of the event with the modes in the factorization theorem illustrated.
\label{fig:factfig}}
\end{center}
\end{figure}

With this understanding of the physical configuration we wish to describe, the proposed factorization theorem for the cusp region of the soft drop groomed heavy hemisphere mass with a single resolved emission in $e^+e^-\to$ hadrons events is
\begin{align}
\frac{d\sigma^{\text{cusp}}}{d\rho}&=2H(Q^2)\int dk^+\, dk^-\,d\rho_q\,d\rho_g\,d\rho_s\,\delta\left(\rho - 4k^+-\rho_q-\rho_g-\rho_s \right)\,\Theta_\text{SD}\\
&
\hspace{1cm}
\times\,H_s(k^+,k^-,\zcut)\,J_q(\rho_q)\, J_g(\rho_g)\,S_{q\bar q g}(\rho_s)\int  d\tau_c\, d\tau_{sc}\, \Theta(\rho-\tau_c-\tau_{sc})\, J_{\bar q}(\tau_c)S_C(\tau_{sc})\,.\nonumber
\end{align}
Here, we express the soft gluon's energy in dimensionless light-cone coordinates with respect to the center-of-mass energy $Q$.  That is, its energy is expressed as
\begin{equation}
k^0 = Q\frac{k^-+k^+}{2}\,,
\end{equation}
for example.  $\Theta_\text{SD}$ are the soft drop groomer phase space constraints from \Eq{eq:sdpsconsts} that ensure that the resolved gluon passes the groomer and that the phase space region where it is collinear to the quark is removed.  The physical picture of the pieces in this factorization theorem is shown in \Fig{fig:factfig}.  In the right hemisphere, its mass is comparable to the grooming parameter, $\rho \sim \zcut$, while in the left hemisphere its mass $\tau$ is parametrically smaller than the grooming parameter, $\tau\ll\zcut$.  

The functions that appear in the factorization theorem are illustrated on a Lund-like plane \cite{Andersson:1988gp} and in physical space in \Fig{fig:factfig}.  On the Lund plane illustration, we have also placed the phase space constraints from grooming and measuring the hemisphere mass and note that functions of the factorization theorem are required at every vertex on that figure, which is a general property for consistency of the factorization theorem.  The various functions in the factorization theorem are:
\begin{enumerate}

\item The hard function $H(Q^2)$ describes the production of the high-energy quark and anti-quark in color-singlet collisions.  At one-loop its expression is \cite{Bauer:2003di,Manohar:2003vb,Ellis:2010rwa,Bauer:2011uc}
\begin{equation}
H(Q^2)=1+\frac{\alpha_sC_F}{4\pi}\left(
-2\,\log^2\frac{\mu^2}{Q^2}-6\log\frac{\mu^2}{Q^2}-16+\frac{7}{3}\pi^2
\right)\,.
\end{equation}

\item The quark or anti-quark jet functions $J_q$ and $J_{\bar q}$ are the jet mass or thrust jet functions because grooming does not affect hard collinear emissions.  To one-loop with the Laplace-conjugate mass $\tilde \rho$, the quark jet function is
\begin{equation}
\tilde J_q(\tilde\rho) =1+\frac{\alpha_s C_F}{4\pi}\left(
2\,\log^2\frac{4\mu^2 \tilde\rho}{Q^2}+3\,\log\frac{4\mu^2 \tilde\rho}{Q^2}+7-\frac{2\pi^2}{3}
\right)\,.
\end{equation}

\item The soft-dropped collinear-soft function $S_C$ describes soft and collinear radiation that both contributes to the hemisphere mass and is affected by the groomer.  At one-loop in Laplace space its expression is \cite{Frye:2016aiz}
\begin{equation}
\tilde S_C(\tilde\rho) = 1 + \frac{\alpha_s\,C_F}{2\pi} \left[-\frac{2+\beta}{2(1+\beta)} \log^2\frac{4\mu^2 \tilde\rho^{2\frac{1+\beta}{2+\beta}}}{Q^2 \, \zcut^{\frac{2}{2+\beta}}} - \frac{\pi^2}{12} \frac{\beta(4+3\,\beta)}{(2+\beta)(1+\beta)} \right],
\end{equation}

\item The hard function for emission of the resolved gluon $H_s(k^+,k^-,\zcut)$ describes both the resolved gluon and wide-angle emissions that fail the groomer.  Its one-loop anomalous dimensions are calculated in \App{app:hardfunc}.

\item The soft function $S_{q\bar q g}$ describes emissions that lie in the angular region between the resolved gluon and the quark.  Its one-loop anomalous dimensions are calculated in \App{app:softfunc}.

\item The resolved gluon's jet function $J_g$ is essentially the same gluon jet function for jet mass, but the scale at which it is evaluated is affected by the form of the measurement here.  The particular scale it is sensitive to is derived in \App{app:gluonjet} and its expression to one-loop in Laplace space is
\begin{equation}
\tilde J_g(\tilde\rho) = 1+ \frac{\alpha_s}{2\pi}  \left[ C_A \log^2\frac{4\mu^2 \tilde\rho}{(k^++k^-)Q^2} + \frac{\beta_0}{2} \, \log\frac{4\mu^2 \tilde\rho}{(k^++k^-)Q^2}+C_A\left(
\frac{67}{18}-\frac{\pi^2}{3}
\right)-\frac{10}{9}n_f T_R
\right]\,.
\end{equation}

\end{enumerate}
As a demonstration of consistency of the factorization theorem, the anomalous dimensions of all functions explicitly sum to 0, which is summarized in \App{app:anomdims}.  We will also show that the fixed-order expansion of this factorization theorem reproduces results for the cusp at leading- and next-to-leading order in $\alpha_s$.

\subsection{Non-Perturbative Corrections}

The form of the convolution in the factorization theorem for the cusp region is slightly different than what is usually present in factorization theorems for other additive observables, like thrust \cite{Schwartz:2007ib}.  In more familiar cases, the value of the observable is set by the sum of contributions from collinear and soft emissions, and no emissions means that the observable value is 0.  Resummation is responsible for pushing the observable distribution away from 0.  However, focusing on the cusp of the soft drop groomer requires that there is a resolved emission that already pushes the observable value away from 0, so that $\rho\sim \zcut$.  Then, about that value, there are additional strongly-ordered soft and collinear emissions that generate an observable value that differs from simply $\rho = 4k^+$.  In the factorization theorem, we then sum over all possible configurations of emissions off of the resolved gluon for which their total sum is the observed value $\rho$.

For the jet and soft functions in the factorization theorem, this sum extends to arbitrarily low transverse momentum emissions; in particular, to emissions that are not described perturbatively.  Thus, at every value of $\rho$, the integral of the factorization theorem includes contributions from non-perturbative emissions.  To account for this non-perturbative contribution, a dispersive approach could be used \cite{Dokshitzer:1995zt}, in which the integral is divided into a perturbative and a non-perturbative regime.  Then, an effective, non-perturbative value of the coupling is used to model the contribution from the non-perturbative region, whose value can be fit to data.  We note that the transverse momentum of a soft, wide-angle non-perturbative emission will be of order of the QCD scale $\Lambda_\text{QCD}\sim$ 1 GeV, and so the non-perturbation contribution to the hemisphere mass will be on the order of
\begin{equation}
\rho_\text{n-p} \sim \frac{\Lambda_\text{QCD}}{Q}\,.
\end{equation}
The grooming parameter is typically taken to be $\zcut \sim 0.1$ experimentally, so at a center-of-mass collision energy of the $Z$ pole (or higher), non-perturbative effects can shift the distribution by $\rho_\text{n-p}\sim 0.01$, or so.  While much more work could go into determining the form of these non-perturbative emissions, like along the lines of \Ref{Hoang:2019ceu}, we leave that for the future.  Here, we will simply freeze the running of $\alpha_s$ at $\mu = 1$ GeV to ensure that our results are sensible, and focus on validating the perturbative resummation as described by the factorization theorem.

\section{Fixed-Order Expansion}\label{sec:foexp}

With the factorization theorem for the cusp region of the groomer in hand, we can then systematically expand it in powers of $\alpha_s$ for validation.  Here, we will show explicitly that it correctly produces the expression for the cusp region at ${\cal O}(\alpha_s)$, and then demonstrate how it captures the dominant terms in the $\rho\sim\zcut\ll 1$ limit at ${\cal O}(\alpha_s^2)$.  For this section to be self-contained, we repeat the factorization theorem for the cusp region here:
\begin{align}
\frac{d\sigma^{\text{cusp}}}{d\rho}&=2H(Q^2)\int dk^+\, dk^-\,d\rho_q\,d\rho_g\,d\rho_s\,\delta\left(\rho - 4k^+-\rho_q-\rho_g-\rho_s \right) \,\Theta_\text{SD}\\
&
\hspace{1cm}
\times\,H_s(k^+,k^-,\zcut)\,J_q(\rho_q)\, J_g(\rho_g)\,S_{q\bar q g}(\rho_s)\int  d\tau_c\, d\tau_{sc}\, \Theta(\rho-\tau_c-\tau_{sc})\, J_{\bar q}(\tau_c)S_c(\tau_{sc})\,.\nonumber
\end{align}

\subsection{Leading Order}

To determine the prediction of the factorization theorem at leading order, we note that the soft gluon emission function $H_s$ starts at ${\cal O}(\alpha_s)$:
\begin{equation}
H_s^{(0)}(k^+,k^-) = \frac{\alpha_s}{\pi}C_F \frac{1}{k^+k^-}\,.
\end{equation}
Thus, to predict the distribution to ${\cal O}(\alpha_s)$, we set all other functions in the factorization theorem to their tree-level expressions.  Thus, we have
\begin{align}\label{eq:locuspcalc}
\frac{d\sigma^{(0),\rho\sim\zcut\ll1}}{d\rho}&=2\int dk^+\, dk^-\,\delta\left(\rho - 4k^+ \right)\,\Theta_\text{SD}\, H_s^{(0)}(k^+,k^-) \,.
\end{align}
Restricting to mMDT or soft drop grooming with $\beta = 0$, we can perform the integrals:
\begin{align}\label{eq:locuspresult}
\frac{d\sigma^{(0),\rho\sim\zcut\ll1}}{d\rho}&=2\frac{\alpha_s C_F}{\pi}\frac{1}{\rho}\left[\Theta\left(2\zcut-\rho \right)\int_{\zcut - \frac{\rho}{4} }^{\zcut}  \frac{dk^-}{k^-}+\Theta\left(\rho-2\zcut \right)\int_{\frac{\rho}{4} }^{\zcut}  \frac{dk^-}{k^-}\right]\\
&=-2\frac{\alpha_s C_F}{\pi}\frac{1}{\rho}\left[\Theta\left(2\zcut-\rho \right)\log\left(
1-\frac{\rho}{4\zcut}
\right)+\Theta\left(\rho-2\zcut \right)\log\frac{\rho}{4\zcut}\right]
\nonumber\,.
\end{align}

To describe the entire regime in which $\rho,\zcut\ll 1$, this can be combined with the result in the hierarchical limit $\rho\ll\zcut\ll 1$, where 
\begin{equation}
\frac{d\sigma^{(0),\rho\ll\zcut\ll 1}}{d\rho} = -2\frac{\alpha_s C_F}{\pi}\frac{1}{\rho}\left[\frac{3}{4}+ \log\,\zcut\right]\,.
\end{equation}
The complete differential cross section in the small $\rho$ and $\zcut$ limit is then
\begin{align}
\frac{d\sigma^{(0),\rho,\zcut\ll 1}}{d\rho}&=\frac{d\sigma^{(0),\rho\sim\zcut\ll 1}}{d\rho}+\frac{d\sigma^{(0),\rho\ll\zcut\ll 1}}{d\rho}\\
&=-2\frac{\alpha_s C_F}{\pi}\frac{1}{\rho}\left[\frac{3}{4}+\Theta\left(2\zcut-\rho \right)\log\left(
\zcut-\frac{\rho}{4}
\right)+\Theta\left(\rho-2\zcut \right)\log\frac{\rho}{4}\right]\,.
\nonumber
\end{align}
This agrees with the explicit expansion of the leading-order cross section from \Ref{Larkoski:2020wgx}.

\subsection{Next-To-Leading Order}

At next-to-leading order, the other functions in the factorization theorem can contribute non-trivially.  The contribution to the differential cross section at ${\cal O}(\alpha_s^2)$ is
\begin{align}
\frac{d\sigma^{(1),{\text{cusp}}}}{d\rho}&=2\int dk^+\, dk^-\,d\rho'\,\delta\left(\rho - 4k^+-\rho' \right)\,H_s^{(0)}(k^+,k^-)\,\Theta_\text{SD} \\
&
\hspace{1cm}
\times\left[
\delta(\rho')\,H^{(1)}(Q^2)+\delta(\rho')\,\frac{H_s^{(1)}(k^+,k^-,\zcut)}{H_s^{(0)}(k^+,k^-)}+J_q^{(1)}(\rho')+ J_g^{(1)}(\rho')+S_{q\bar q g}^{(1)}(\rho')\right.\nonumber\\
&
\hspace{8cm}
\left.+\, \delta(\rho')\int_0^\rho  d\tau\,\left( J_{\bar q}^{(1)}(\tau)+S_c^{(1)}(\tau)\right)
\right]\nonumber\,.
\end{align}
All of the one-loop expressions for the functions in the factorization theorem were presented in \Sec{sec:fact} or are calculated in the appendices.  While in principle everything is known to completely evaluate these integrals, they quickly become extremely unwieldy.  As a result, we will restrict to the calculation of the contribution at ${\cal O}(\alpha_s^2)$ that is proportional to the number of active quarks, $n_f$.  Also, as earlier, we will restrict to the mMDT groomer for simplicity, but results for soft drop with any value of $\beta$ can be obtained using the results in the appendices.

\subsubsection{$n_f$ Channel}

For NLL resummation, the only contributions to the $n_f$ color channel arise from terms proportional to the $\beta$-function.  The one-loop coefficient of the $\beta$-function is
\begin{equation}
\beta_0 = \frac{11}{3}C_A - \frac{4}{3}T_R n_f\,.
\end{equation}
The only functions in the factorization theorem that have $\beta$-function dependence are the resolved gluon emission function and the gluon's jet function for which we have
\begin{align}
\frac{H_s^{(1)}(k^+,k^-,\zcut)}{H_s^{(0)}(k^+,k^-)} &\supset-\frac{\alpha_s}{4\pi}\beta_0\log\frac{\mu^2}{Q^2}\,,\\
J_g^{(1)}(\rho') &\supset \frac{\alpha_s }{2\pi}\delta(\rho')\frac{\beta_0}{2}\log\frac{4\mu^2}{(k^++k^-)Q^2}-\frac{\alpha_s }{\pi}\frac{\beta_0}{4}\left(\frac{1}{\rho'}\right)_+\,,
\end{align}
which are the expressions relevant for NLL accuracy.  From the form of the logarithm in the resolved gluon emission function $H_s$, we evaluate the coupling at the center-of-mass collision energy, $\alpha_s \equiv \alpha_s(Q)$.  This will enable the simplest numerical comparison to EVENT2 later.  The contribution to the cross section in this color channel to this order is therefore 
\begin{align}
&\frac{d\sigma^{(1),n_f}}{d\rho} \\
&\hspace{0.5cm}= \frac{8}{3}\left(\frac{\alpha_s }{2\pi}\right)^2C_FT_Rn_f\int \frac{dk^+}{k^+}\, \frac{dk^-}{k^-}\, d\rho'\,\delta(\rho-4k^+ - \rho')\,\Theta_\text{SD}
\left(
\left(
\frac{1}{\rho'}
\right)_+ - \delta(\rho')\log\frac{4}{k^++k^-}
\right)
\nonumber\,.
\end{align}

The term with the $\delta(\rho')$ factor is
\begin{align}
&-\int \frac{dk^+}{k^+}\, \frac{dk^-}{k^-}\, d\rho'\,\delta(\rho-4k^+ - \rho')\,\Theta_\text{SD}\,
 \delta(\rho')\,\log\frac{4}{k^++k^-}
\\
&\hspace{1cm}
=-\frac{1}{\rho}\left[
-\frac{\pi^2}{6}-\frac{1}{2}\log^2\left(
\zcut-\frac{\rho}{4}
\right)-\log\left(
\zcut-\frac{\rho}{4}
\right)\log\frac{4}{\rho}-\frac{1}{2}\log^2\frac{4}{\rho}-\frac{1}{2}\log^2\frac{4}{\zcut}\right.\nonumber\\
&
\hspace{3cm}
\left.
+\frac{1}{2}\log^2\left(
\frac{4}{\zcut-\frac{\rho}{4}}
\right)-\text{Li}_2\left(
1-\frac{4\zcut}{\rho}
\right)
\right]\Theta(2\zcut-\rho)\nonumber\\
&
\hspace{1.5cm}-\frac{1}{\rho}\left[
-\frac{\pi^2}{12}-\frac{1}{2}\log^2\frac{4}{\zcut}+\frac{1}{2}\log^2\frac{16}{\rho}
\right]\Theta(\rho-2\zcut)
\nonumber\,.
\end{align}
Recall that the prescription for the phase space constraints imposed by $\Theta_\text{SD}$ is defined in \Eq{eq:psconstsprescrip}.  The integral with the $+$-function can be rearranged according to the bounds imposed by the grooming phase space constraints, $\Theta_\text{SD}$:
\begin{align}
&\int  \frac{dk^-}{k^-}\, \frac{d\rho'}{\rho-\rho'}\,\Theta_\text{SD}\, \left(
\frac{1}{\rho'}
\right)_+\\
&\hspace{1cm}
=\int  \frac{d\rho'}{\rho-\rho'}\,\left(
\frac{1}{\rho'}
\right)_+\,\left[
-\log\left(
1-\frac{\rho-\rho'}{4\zcut}
\right)\,\Theta(\rho-\rho')\,\Theta\left(
\rho'-(\rho-2\zcut)
\right)\right.\nonumber\\
&
\hspace{9cm}
\left.
+\,\log\frac{4\zcut}{\rho-\rho'}\, \Theta\left((\rho-2\zcut) - \rho'\right)
\right]\nonumber\\
&
\hspace{1cm}
=\int\frac{d\rho'}{\rho-\rho'}\, \frac{1}{\rho'}\left[
-\log\left(
1-\frac{\rho-\rho'}{4\zcut}
\right)\,\Theta(\rho-\rho')\,\Theta\left(
\rho'-(\rho-2\zcut)
\right)\,\Theta(\rho-2\zcut)
\right]\nonumber\\
&
\hspace{1cm}
+\int_0  \frac{d\rho'}{\rho-\rho'}\,\left(
\frac{1}{\rho'}
\right)_+\,\left[
-\log\left(
1-\frac{\rho-\rho'}{4\zcut}
\right)\,\Theta(\rho-\rho')\,\Theta\left(
2\zcut-\rho
\right)\right.\nonumber\\
&
\hspace{7cm}
\left.
+\,\log\frac{4\zcut}{\rho-\rho'}\, \Theta\left((\rho-2\zcut) - \rho'\right)\,\Theta\left(
\rho-2\zcut
\right)
\right]
\nonumber
\end{align}
The first integral with no $+$-function is
\begin{align}
&-\int\frac{d\rho'}{\rho-\rho'}\, \frac{1}{\rho'}
\log\left(
1-\frac{\rho-\rho'}{4\zcut}
\right)\,\Theta(\rho-\rho')\,\Theta\left(
\rho'-(\rho-2\zcut)
\right)\,\Theta(\rho-2\zcut)\\
&
\hspace{1cm}
=\frac{1}{\rho}\left[\frac{\pi^2}{12} + \log\, 2\, \log \frac{2\zcut}{\rho-2\zcut}+\text{Li}_2\left(
1-\frac{\rho}{4\zcut}
\right)-\text{Li}_2\left(
2-\frac{\rho}{2\zcut}
\right)\right]\Theta(\rho-2\zcut)
\nonumber\,.
\end{align}

To continue and evaluate the integrals with explicit $+$-functions, use the definition of the $+$-function, where
\begin{equation}
\int_0^1d\rho' \,\left(\frac{1}{\rho'}\right)_+ = 0\,,
\end{equation}
and so
\begin{equation}
\left(\frac{1}{\rho'}\right)_+=\lim_{\epsilon \to 0}\left[
\frac{1}{\rho'}\,\Theta(\rho'-\epsilon) + \log\,\epsilon \,\delta(\rho'-\epsilon)
\right]\,.
\end{equation}
Using this, the first integral with the $+$-function is
\begin{align}
&-\int_0  \frac{d\rho'}{\rho-\rho'}\,\left(
\frac{1}{\rho'}
\right)_+\,
\log\left(
1-\frac{\rho-\rho'}{4\zcut}
\right)\,\Theta(\rho-\rho')\,\Theta\left(
2\zcut-\rho
\right)\\
&
\hspace{1cm}
=\frac{1}{\rho}\left[
-\frac{\pi^2}{6}+\frac{\log^2\left(
1-\frac{\rho}{4\zcut}
\right)}{2}-\log(4\zcut-\rho)\log\left(
1-\frac{\rho}{4\zcut}
\right)\right.\nonumber\\
&
\hspace{6cm}\left.
+\, \text{Li}_2\left(
\frac{\rho}{4\zcut}
\right)+\text{Li}_2\left(
1-\frac{\rho}{4\zcut}
\right)
\right]\Theta(2\zcut-\rho)
\nonumber\,.
\end{align}
The second integral with the $+$-function is
\begin{align}
&\int_0  \frac{d\rho'}{\rho-\rho'}\,\left(\frac{1}{\rho'}\right)_+\,\log\frac{4\zcut}{\rho-\rho'}\, \Theta\left((\rho-2\zcut) - \rho'\right)\,\Theta\left(
\rho-2\zcut
\right)\\
&
\hspace{1cm}
=\frac{1}{\rho}\left[
\frac{\pi^2}{6}+\frac{\log^2 2}{2}+\log\, 2\, \log\left(1-\frac{2\zcut}{\rho}\right)-\frac{\log^2\frac{4\zcut}{\rho}}{2}+\log\frac{4\zcut}{\rho}\,\log\, \rho\right.\nonumber\\
&
\hspace{9cm}
\left. - \text{Li}_2\left(
\frac{2\zcut}{\rho}
\right)
\right]\Theta\left(
\rho-2\zcut
\right)
\nonumber\,.
\end{align}

Putting it all together, the complete contribution at ${\cal O}(\alpha_s^2)$ for the $n_f$ channel is: 
\begin{align}\label{eq:nfnlopred}
&\left(
\frac{2\pi}{\alpha_s}
\right)^2\frac{\rho}{C_F  T_R n_f}\frac{d\sigma^{(1),n_f}}{d\rho} \\
&
\hspace{1cm}=\frac{8}{3}\left[
\frac{1}{2}\log^2\left(
\zcut-\frac{\rho}{4}
\right)+\log\left(
\zcut-\frac{\rho}{4}
\right)\log\frac{4}{\rho}+\frac{1}{2}\log^2\frac{4}{\rho}+\frac{1}{2}\log^2\frac{4}{\zcut}-\frac{1}{2}\log^2\left(
\frac{4}{\zcut-\frac{\rho}{4}}
\right)\right.\nonumber\\
&
\hspace{3cm}+\frac{1}{2}\log^2\left(
1-\frac{\rho}{4\zcut}
\right)-\log(4\zcut-\rho)\log\left(
1-\frac{\rho}{4\zcut}
\right)
\nonumber\\
&\hspace{3cm}
\left.
+\,\text{Li}_2\left(
1-\frac{4\zcut}{\rho}
\right)+\text{Li}_2\left(
\frac{\rho}{4\zcut}
\right)+\text{Li}_2\left(
1-\frac{\rho}{4\zcut}
\right)
\right]\Theta(2\zcut-\rho)\nonumber\\
&
\hspace{2cm}
+\frac{8}{3}\left[\frac{\pi^2}{3}+
\log\frac{\rho^2}{8}\,\log\,\zcut-\frac{1}{8}\log^2\frac{128}{\rho^4}-\frac{3}{8}\log^22+\text{Li}_2\left(
1-\frac{\rho}{4\zcut}
\right)-\text{Li}_2\left(
2-\frac{\rho}{2\zcut}
\right)\right.\nonumber\\
&
\hspace{3cm}\left.-\,\text{Li}_2\left(\frac{2\zcut}{\rho}
\right)
\right]
\Theta(\rho-2\zcut)\,.\nonumber
\end{align}

\begin{figure}
\begin{center}
\includegraphics[width=7cm]{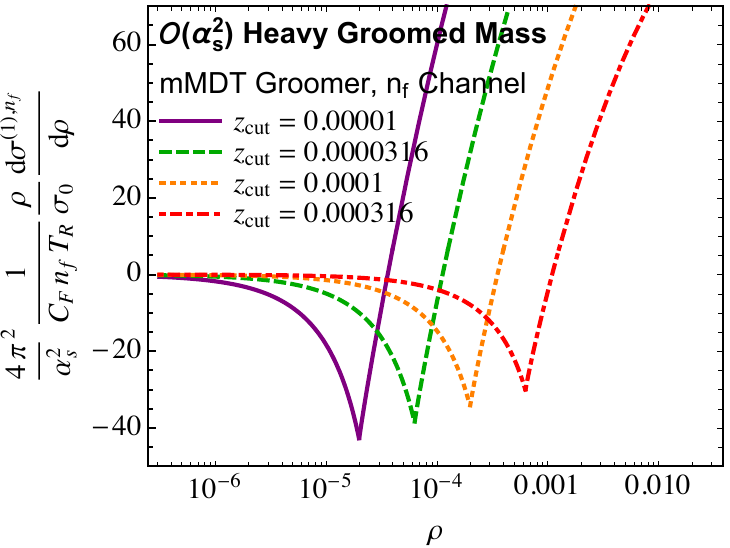}
\hspace{0.5cm}
\includegraphics[width=7cm]{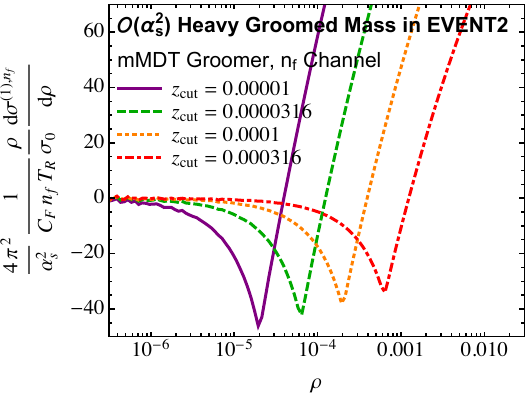}
\caption{
Comparison between the prediction from \Eq{eq:nfnlopred} (left) and EVENT2 (right) of the cusp region of the mMDT groomed heavy hemisphere mass in the $n_f$ channel at next-to-leading order for various values of $\zcut$.  The contribution in the $\rho\ll \zcut\ll 1$ limit has been removed.
\label{fig:nfchannel}
}
\end{center}
\end{figure}

\begin{figure}
\begin{center}
\includegraphics[width=7cm]{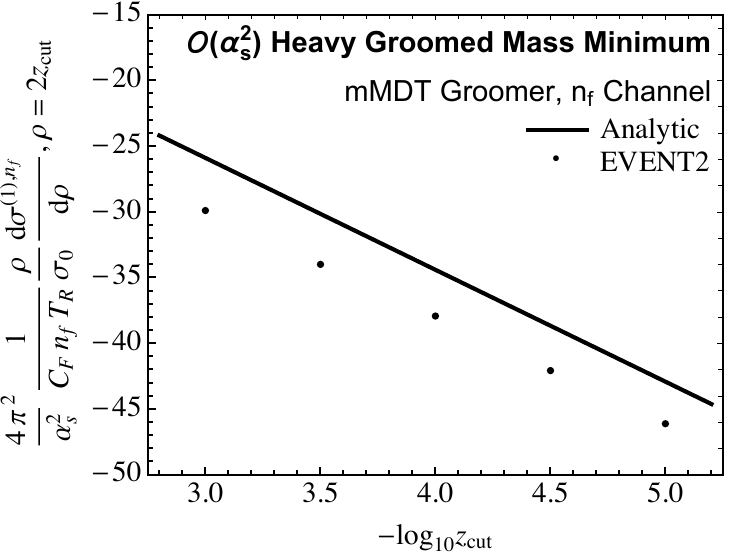}
\caption{
Comparison between the value of the $n_f$ channel of the ${\cal O}(\alpha_s^2)$ contribution to the differential cross section for the mMDT groomed heavy hemisphere mass at the cusp location of $\rho=2\zcut$ from \Eq{eq:cuspdiffxsecval} and EVENT2, as a function of $\zcut$.  
\label{fig:nfmin}
}
\end{center}
\end{figure}

This analytic expression for the cusp region can be directly compared to numerical output of $e^+e^-\to q\bar q g$ events at next-to-leading order from EVENT2 \cite{Catani:1996vz}.  We generated about $10^{13}$ events with $\zcut$ values ranging down to $10^{-5}$ and measured the groomed heavy hemisphere mass.  Then, from the complete differential cross section, we subtract the analytic expression for the $\rho\ll\zcut\ll 1$ regime, whose expression can be found in \Ref{Kardos:2020ppl}.  The resulting distributions then only describe the cusp region $\rho\sim \zcut$, and larger values of $\rho$ and can be directly compared to the prediction of \Eq{eq:nfnlopred}.  The results of this comparison restricted to the $n_f$ channel are shown in \Fig{fig:nfchannel}.  Excellent qualitative agreement between our analytic calculation and EVENT2 is manifest in the cusp region around $\rho = 2\zcut$, for various values of $\zcut$.  In particular, the trend that the cross section value at the cusp decreases as $\zcut$ decreases is exhibited in the analytic expression and EVENT2.  This can be directly compared.  At the cusp, where $\rho =2\zcut$, the predicted value of the differential cross section is 
\begin{align}\label{eq:cuspdiffxsecval}
\left.\left(
\frac{2\pi}{\alpha_s}
\right)^2\frac{\rho}{C_F  T_R n_f}\frac{d\sigma^{(1),n_f}}{d\rho}\right|_{\rho=2\zcut} &=\frac{2\pi^2}{9}-\frac{16}{3}\log^2 2+\frac{16}{3}\log\, 2\, \log\,\zcut\\
&
\simeq -0.3692+3.6968\,\log\,\zcut
\,.\nonumber
\end{align}
One can also extract the value from EVENT2, and the plot comparing them is shown in \Fig{fig:nfmin}.  The slope of the prediction and EVENT2 agree extremely well, though there is a constant offset.  This is to be expected: there are terms at ${\cal O}(\alpha_s^2)$ in the $n_f$ channel that have not been included in the factorization theorem, as they formally lie beyond NLL accuracy.  However, the dominant contribution has been captured in the factorization theorem to this accuracy, and the residual difference is relatively small.

\section{Resummation}\label{sec:resum}

Having validated our factorization theorem for the cusp region of the soft drop groomer, we now present the expression for the groomed heavy hemisphere cross section at next-to-leading logarithmic accuracy (NLL), following the procedure explicitly presented in Appendix F of \Ref{Frye:2016aiz}.  For compactness, we will only explicitly show the expression for soft drop with $\beta = 0$ (mMDT grooming), but results for arbitrary grooming exponent are presented in the appendices.  We have  
\begin{align}\label{eq:resumnllfinal}
		&\frac{d\sigma^{\text{cusp,NLL}}}{d\rho}\\
		 &\hspace{0cm}= 2\int dk^+\, dk^-\,\Theta_\text{SD}\,\Theta(\rho-4k^+)\,\frac{\alpha_s(\mu_{H_s})}{\pi} \frac{C_F}{k^+ k^-}\frac{1}{\rho-4k^+} \nonumber\\
		 &\hspace{0.5cm}\times \left[
		 \Theta(\mu_{H_s}-\mu_S)+\Theta(\mu_S-\mu_{H_s})\left(
1-\frac{k^+}{k^-}
\right)^{\frac{2C_A}{\beta_0}\log\frac{\alpha_s(\mu_{H_s})}{\alpha_s(\mu_S)}}
		 \right]\nonumber\\ 
			&\hspace{0.5cm}\times\exp\left[K_H(\mu, \mu_H) + K_{J_L}(\mu, \mu_{J_L}) + K_{J_R}(\mu, \mu_{J_R}) + K_{S_C}(\mu, \mu_{S_C})+ K_{J_g}(\mu, \mu_{J_g}) \right.\nonumber\\
			&
			\hspace{10.5cm}
			\left.+ K_{H_s}(\mu, \mu_{H_s}) + K_S(\mu, \mu_S) \right] \nonumber\\
		&\hspace{0.5cm}\times \left[1 + \frac{\alpha_s(\mu_H) C_F}{4\pi}\left(-2\log^2\frac{\mu_H^2}{Q^2} - 6\log\frac{\mu_H^2}{Q^2} \right)+ \frac{\alpha_s(\mu_{J_L}) C_F}{4\pi}\left(2\partial^2_{\omega_{J_L}} + 3\partial_{\omega_{J_L}} \right)\phantom{\frac{4C_A I\left(
		\frac{k^+}{k^-}
		\right)}{\pi}}\right. \nonumber\\
		&\hspace{1cm}+ \frac{\alpha_s(\mu_{J_R}) C_F}{4\pi}\left(2\partial^2_{\omega_{J_R}} + 3\partial_{\omega_{J_R}} \right) - \frac{\alpha_s(\mu_{S_C}) C_F}{2\pi} \,\partial_{\omega_{S_C}}^2+ \frac{\alpha_s(\mu_{J_g})}{2\pi}\left(C_A \partial^2_{\omega_{J_g}} + \frac{\beta_0}{2} \, \partial_{\omega_{J_g}}\right)   \nonumber\\
		&\hspace{1cm}+ \frac{\alpha_s(\mu_{H_s})}{4\pi}\left((C_F- C_A)\log^2 \frac{(k^+k^-)^{\frac{C_A}{C_F-C_A}}\mu_{H_s}^2 }{Q^2  \zcut^{\frac{2C_F}{C_F-C_A}}}\right.\nonumber\\
			&\hspace{3.5cm} \left.- \left(\beta_0 + \frac{4 C_F A\left(
		\frac{k^+}{k^-}
		\right)}{\pi} + \frac{4 C_A I\left(
		\frac{k^+}{k^-}
		\right)}{\pi}  \right) \log \frac{(k^+k^-)^{\frac{C_A}{C_F-C_A}}\mu_{H_s}^2 }{Q^2  \zcut^{\frac{2C_F}{C_F-C_A}}} \right)  \nonumber\\
		&\hspace{1cm}\left.+ \frac{\alpha_s(\mu_S)}{4\pi}\left(-(C_F + C_A)\partial^2_{\omega_S} + \left(\frac{4C_F A\left(
		\frac{k^+}{k^-}
		\right)}{\pi} + \frac{4C_A I\left(
		\frac{k^+}{k^-}
		\right)}{\pi}- 2 C_A \log\frac{k^+ k^-}{(k^- + k^+)^2}\right)\partial_{\omega_S} \right)\right] \nonumber\\
		&\hspace{0.5cm}\times \left(\frac{\mu_H^2}{Q^2}\right)^{\omega_H(\mu, \mu_H)} 
		\left(\frac{4\mu_{J_L}^2}{Q^2 \rho}\right)^{\omega_{J_L}(\mu, \mu_{J_L})} 
		\left(\frac{4\mu_{S_C}^2}{Q^2 \zcut \rho}\right)^{\omega_{S_C}(\mu, \mu_{S_C})}\left(\frac{4\mu_{J_R}^2}{Q^2}\frac{1}{\rho - 4k^+}\right)^{\omega_{J_R}(\mu, \mu_{J_R})} \nonumber\\ 
		&\hspace{0.5cm}\times 
		\left(\frac{4\mu_{J_g}^2}{(k^++k^-)Q^2} \frac{1}{\rho - 4k^+}\right)^{\omega_{J_g}(\mu, \mu_{J_g})} \left(\frac{(k^+k^-)^{\frac{C_A}{C_F-C_A}}\mu_{H_s}^2 }{Q^2  \zcut^{\frac{2C_F}{C_F-C_A}}}\right)^{\omega_{H_s}(\mu, \mu_{H_s})}
		\left(\frac{16\mu_S^2}{Q^2}\frac{1}{(\rho - 4k^+)^2}\right)^{\omega_S(\mu, \mu_S)} \nonumber\\
		&\hspace{0.5cm}\times \Gamma\left(1-\omega_{J_L}(\mu, \mu_{J_L}) - \omega_{S_C}(\mu, \mu_{S_C})\right)^{-1} \Gamma\left(-\omega_{J_R}(\mu, \mu_{J_R}) - \omega_{J_g}(\mu, \mu_{J_g}) - 2\omega_S(\mu, \mu_S)\right)^{-1} . \nonumber
\end{align}
In this expression, the exponential factors for function $F$ expanded to NLL order are
\begin{align}\label{eq:NLL K function}
\!\! K_F(\mu,\mu_0)&=C_i\frac{\Gamma_0}{2\beta_0^2}\left[
\frac{4\pi}{\alpha_s(\mu_0)}\left(
\log \,r+\frac{1}{r}-1
\right)+\left(
\frac{\Gamma_1}{\Gamma_0}-\frac{\beta_1}{\beta_0}
\right)(r-1-\log\, r) \right]-\frac{\gamma_0}{2\beta_0}\log\, r\,,
\end{align}
where $$r=\frac{\alpha_s(\mu)}{\alpha_s(\mu_0)}\,.$$  The other exponential factors, again to NLL accuracy, are
\begin{align}\label{eq:NLL omega function}
\omega_F(\mu,\mu_0)&=-C_i\frac{\Gamma_0}{{ 2 }\beta_0}\left[
\log\,r+\frac{\alpha_s(\mu_0)}{4\pi}\left(
\frac{\Gamma_1}{\Gamma_0}-\frac{\beta_1}{\beta_0}
\right)(r-1)
\right] \,.
\end{align}
In both of these expressions, the $\beta$-function, cusp anomalous dimension, and non-cusp anomalous dimension are expressed as a series in $\alpha_s$ and the coefficients of their series appear in these factors.  Specifically, the $\beta$-function is
\begin{equation}
\beta(\alpha_s)=\mu\frac{\partial\alpha_s}{\partial\mu} =-2\alpha_s\sum_{n=0}^\infty \beta_n \left(
\frac{\alpha_s}{4\pi}
\right)^{n+1}\,.
\end{equation}
For NLL resummation, we need the $\beta$-function to two-loop order \cite{Tarasov:1980au}.  The first two coefficients are
\begin{align}
&
\hspace{-0.5cm}
\beta_0 =\frac{11}{3}C_A -\frac{4}{3}T_R n_f \,, \\
&
\hspace{-0.5cm}\beta_1 =\frac{34}{3}C_A^2-4T_R n_f\left(
C_F+\frac{5}{3}C_A
\right) \,.\nonumber
\end{align}
The cusp anomalous dimension is
\begin{equation}\label{eq:cuspexp}
\Gamma_\text{cusp} = \sum_{n=0}^\infty \Gamma_n \left(
\frac{\alpha_s}{4\pi}
\right)^{n+1}\,.
\end{equation}
The first two coefficients of the cusp anomalous dimension are \cite{Korchemsky:1987wg}:
\begin{align}
\Gamma_0 &= 4 \,,\\
\Gamma_1 &=  4C_A\left(
\frac{67}{9}-\frac{\pi^2}{3}
\right) - \frac{80}{9}T_R n_f
\nonumber \,.
\end{align}
Finally, the non-cusp anomalous dimension is
\begin{equation}
\gamma = \sum_{n=0}^\infty \gamma_n \left(
\frac{\alpha_s}{4\pi}
\right)^{n+1}\,.
\end{equation}
$C_i$ is the appropriate color factor for the function $F$ appearing in the factorization theorem and the non-cusp anomalous dimensions can be extracted from the results of \App{app:anomdims}.

The terms in the large square brackets of \Eq{eq:resumnllfinal} are the low-scale functions, expanded to first-order in $\alpha_s$.  Note that some of these terms have derivatives that act on the appropriate $\omega_F$ factor appearing later in the expression.  In the results we present shortly, we will not perform detailed analysis of scale variations as a measurement of uncertainty, as our emphasis here is to just demonstrate that the factorization theorem describes the cusp.  With that understanding, we will evaluate this resummed expression with canonical scales, for which the scales $\mu_F$ are selected such that the factors raised to the $\omega_F$ powers are unity.  For example, the canonical scale of the soft function $\mu_S$ is
\begin{equation}
\mu_S = \frac{(\rho-4k^+)Q}{4}\,.
\end{equation}
Finally, we choose the renormalization scale $\mu= Q = m_Z$, corresponding to $e^+e^-$ collisions at the $Z$-pole and the choice which sets the hard function $H(Q^2) = 1$.  Additionally, we do not attempt any sophisticated modeling of non-perturbative corrections at low scales.  To avoid the Landau pole of QCD, we freeze the value of $\alpha_s$ when the scale at which it is evaluated falls below 1 GeV.  This prescription does require using expressions for the $K_F$ and $\omega_F$ functions in the resummation result derived with a non-running coupling, as described in \App{app:fixed-coupling}.

With this prescription for making the plots that we show in the following, we also note that for a center-of-mass collision energy at the $Z$-pole, the canonical scale of the soft function is nearly non-perturbative itself.  Around the cusp region, $\rho-4k^+\lesssim\zcut$ and so for typical values of $\zcut \sim 0.1$, the soft canonical scale is bounded above by a couple of GeV.  So, the plots that we will show for this resummation at the cusp are mostly illustrative, rather than phenomenologically relevant.  Nevertheless, potential future higher energy $e^+e^-$ colliders would enable using smaller values for $\zcut$ that are still dominated by perturbative physics, and so this resummation of the cusp could be important.

The factor in the large square brackets on the third line of \Eq{eq:resumnllfinal} corresponds to resummation in the limit where the resolved gluon approaches the hemisphere boundary, when $k^--k^+\ll k^-$.  This factor,
$$
\Theta(\mu_{H_s}-\mu_S)+\Theta(\mu_S-\mu_{H_s})\left(
1-\frac{k^+}{k^-}
\right)^{\frac{2C_A}{\beta_0}\log\frac{\alpha_s(\mu_{H_s})}{\alpha_s(\mu_S)}}
$$
suppresses collinear emissions off of the resolved gluon if they land in the other hemisphere and correspond to a scale larger than that in that other hemisphere.  These emissions and the form of this term is precisely that described by the ``dressed gluon approximation'' and the boundary soft modes of \Ref{Larkoski:2015zka}, which has been used to resum non-global logarithms \cite{Dasgupta:2001sh}.  The form of non-global contributions to the grooming cusp is a bit different than how they arise in, say, just the traditional hemisphere mass, but their effect is essentially identical.  Emissions from the heavier hemisphere into the lighter hemisphere are suppressed if those emissions would have parametrically affected the scale in the lighter hemisphere.  However, in the groomed mass case at hand, the lighter hemisphere is not directly contributing to the value of the observable.  The calculation of this boundary factor from our factorization theorem is presented in \App{app:boundarysofts}.

\subsection{Double-Logarithmic Limit}

This complete NLL-accurate expression is still extremely unwieldy, and we will only be able to evaluate it numerically for making predictions later.  So it is enlightening to expand it out, keeping only those terms that contribute at double-logarithmic accuracy with fixed coupling.  Performing this expansion, the differential cross section about the cusp simplifies to
\begin{align}\label{eq:dllcusp}
&\frac{d\sigma^{\text{cusp,DL}}}{d\rho} \\
&
\hspace{0.5cm}= -2C_F(C_F+C_A)\frac{\alpha_s^2 }{\pi^2}\int \frac{dk^+}{k^+}\,\frac{dk^-}{k^-}\,\frac{\log\left(
\rho-4k^+
\right)}{\rho-4k^+}\,e^{-\frac{\alpha_s}{2\pi}(C_F+C_A)\log^2\left(
\rho-4k^+
\right)}\,\Theta(\rho-4k^+)\Theta_\text{SD}\,.\nonumber
\end{align}
In this form, $k^-$ can be explicitly analytically integrated over, but we won't do that here.  This expression explicitly demonstrates that resummation, even at this limited accuracy, is responsible for smoothing out the $\delta$-function that appears in the leading-order expression for the cross section at the cusp, in \Eq{eq:locuspcalc}.  Specifically, note that in the limit that $\alpha_s\to 0$, the exponential factor of $\rho-4k^+$ reduces to the $\delta$-function:
\begin{align}
\lim_{\alpha_s\to 0}\left[-\frac{\alpha_s}{\pi}(C_F+C_A)\frac{\log\left(
\rho-4k^+
\right)}{\rho-4k^+}\,e^{-\frac{\alpha_s}{2\pi}(C_F+C_A)\log^2\left(
\rho-4k^+
\right)}\right] = \delta(\rho-4k^+)\,.
\end{align}

\begin{figure}
\begin{center}
\includegraphics[width=7cm]{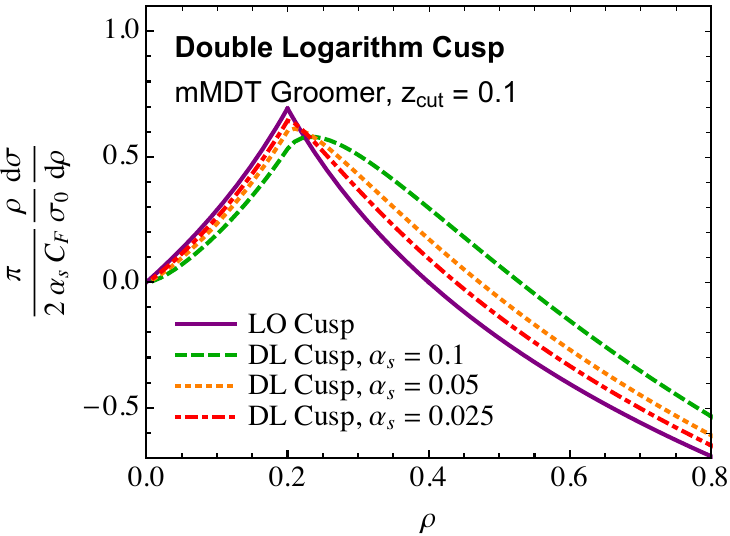}
\hspace{0.5cm}
\includegraphics[width=7cm]{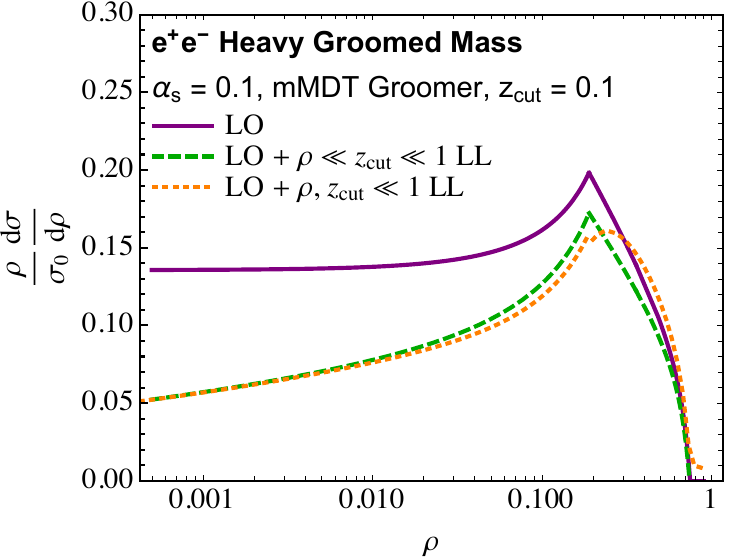}
\caption{
Left: Plots of the double-logarithmic resummed distribution \Eq{eq:dllcusp} about the cusp for various values of $\alpha_s$.  Also plotted is the leading-order distribution from \Eq{eq:locuspresult}.  Right: Comparison of the leading-order (LO), leading-order matched with $\rho\ll\zcut\ll 1$ resummation, and subsequently then matched with $\rho\sim \zcut \ll1$ resummation.
\label{fig:dll_matched}
}
\end{center}
\end{figure}

This limit can be directly observed in the resulting distribution from \Eq{eq:dllcusp}.  On the left of \Fig{fig:dll_matched}, we have plotted the double-logarithmic resummed distribution with different values of $\alpha_s$, and have divided out a factor of $\alpha_s$.  For relatively large values of $\alpha_s$, the cusp is smoothed out and as $\alpha_s$ decreases, the distribution approaches the leading-order result, with a non-smooth cusp at $\rho = 2\zcut$.  Further, with this distribution around the cusp region, we can match it with the leading, fixed-order distribution, as well as with the leading-logarithmic resummed result in the limit where $\rho\ll \zcut\ll 1$.  In this limit, the resummed distribution with fixed-coupling is \cite{Dasgupta:2013ihk,Larkoski:2014wba,Frye:2016aiz}
\begin{align}
\frac{d\sigma^{\rho\ll\zcut\ll1\text{,LL}}}{d\rho} = -2\frac{\alpha_s C_F}{\pi}\frac{1}{\rho}\left(
\log\,\zcut + \frac{3}{4}
\right)\,e^{-2\frac{\alpha_s}{\pi}C_F\left(
\log\,\zcut + \frac{3}{4}
\right)\log\,\rho}\,.
\end{align}
On the right of \Fig{fig:dll_matched}, we show the results of this matching, with the leading-order (LO), LO matched with the $\rho\ll\zcut\ll 1$ resummation, and then that matched with the $\rho\sim\zcut \ll 1$ resummation.  To match, we have simply added the results together and subtracted their overlap, which is possible because all divergences at leading order in the $\rho\ll\zcut\ll 1$ limit are accounted for in the resummation.  Note that the matched result when the cusp resummation is included has a small, residual cusp around $\rho = 2\zcut$.  This occurs because the precise location of the cusp at leading-order is actually at $\rho = 2\zcut - \zcut^2$ \cite{Larkoski:2020wgx}.  As $\zcut$ decreases, this residual cusp also decreases, matching smoothly onto the limiting location of $\rho = 2\zcut$.

\subsection{Matching with $\rho\ll\zcut\ll1$ Factorization and Fixed-Order}

\begin{figure}
\begin{center}
\includegraphics[width=7cm]{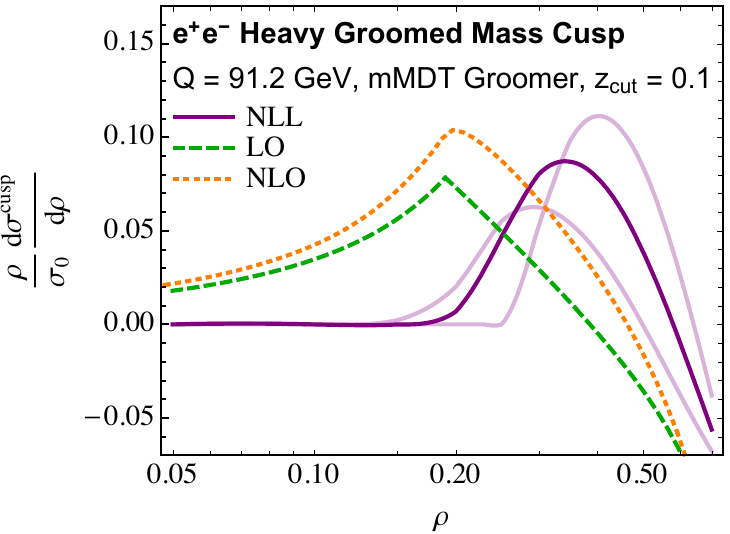}
\hspace{0.5cm}
\includegraphics[width=7cm]{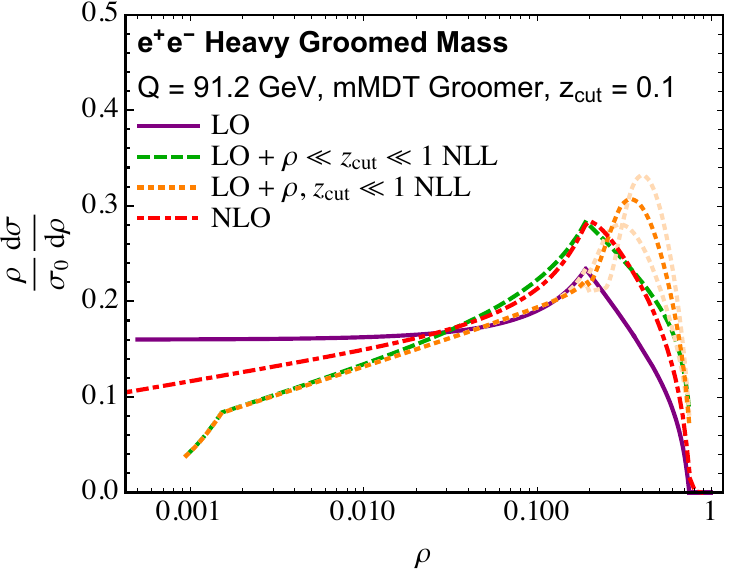}
\caption{
Plots comparing the NLL resummation of the cusp region \Eq{eq:resumnllfinal} to other calculations.  Left: Comparison of the cusp resummation to the cusp at leading- and next-to-leading fixed order.  The darker NLL distribution uses canonical scales, while the lighter curves to the left (right) vary the scale of the soft function up (down) by a factor of 2.  Right: Comparison of the full groomed heavy hemisphere mass distribution at fixed-order and matched to resummed results.  Again, the lighter curves correspond to variation of the soft scale in the NLL cusp resummation up or down by a factor of 2.
\label{fig:lo_matched}
}
\end{center}
\end{figure}

With the complete expression for the resummed result for the groomed heavy hemisphere mass distribution at next-to-leading logarithmic accuracy in hand, we are now in a position to evaluate it and then match to fixed-order and resummed results in other phase space regimes.  To produce numerical results for the distribution of \Eq{eq:resumnllfinal}, we use the {\tt Vegas} integration implementation in {\sc Cuba 4.2} \cite{Hahn:2004fe}.  Here we will just show some representative plots and will not attempt an exhaustive presentation of dependence of the distribution on parameters and collision energy.  With this caveat, we restrict to collisions at the $Z$ pole, $Q = 91.2$ GeV and fix $\zcut = 0.1$.  We first compare the NLL distribution about the cusp to the leading- and next-to-leading fixed order results on the left of \Fig{fig:lo_matched}.  The leading-order distribution was calculated analytically in \Ref{Larkoski:2020wgx}, and the next-to-leading order distribution was calculated with EVENT2 \cite{Catani:1996vz} with the $\rho\ll\zcut\ll 1$ limit subtracted.  For the resummed result, we present three distributions corresponding to canonical scale choices as described earlier, and then distributions whose scale of the soft function is varied up and down by a factor of 2 (lighter on the plot).  The most striking difference between fixed-order and resummed results is the location of the peak, with the resummed results peaking at very large values of $\rho$.  As $\alpha_s$ increases, this cusp distribution moves right to larger values of $\rho$ and, for $Q = 91.2$ GeV and $\zcut = 0.1$, the canonical scale of the soft function is barely perturbative.  So, in these collisions, all-orders resummation of the cusp is likely not relevant.  However, as the collision energy increases, soft emissions that smear the cusp become more and more perturbative, and so resummation may be important for future $e^+e^-$ colliders.  We present plots of the resummation at the cusp in 1 TeV $e^+e^-$ collisions in \App{app:1tevcoll}.

In the plot on the right of \Fig{fig:lo_matched}, we present the complete groomed heavy hemisphere distribution calculated at various accuracies.  On this plot, the leading- and next-to-leading order distributions are presented, as well as the NLL resummed distribution in the $\rho \ll\zcut\ll 1$ limit \cite{Frye:2016aiz} matched to leading-order.  Further, we also match this distribution with our NLL resummation of the cusp region.  For both of these matched distributions, we have simply added the fixed-order and resummed results, and then subtracted their overlap.  The NLL resummation at the cusp only smoothes out the region around the cusp itself, and has no effect on the distribution in the regime where $\rho\ll\zcut$.  As observed earlier, because of the very low scale of the soft function, the cusp resummation pushes this region to much larger values of $\rho$.

\section{Conclusions}\label{sec:concs}

In this paper, we have established an all-orders factorization theorem for groomed observables in the region of phase space where the scale set by the groomer is comparable to the scale set by the observable.  We focused on validation of the factorization theorem as describing strongly-ordered soft and collinear emissions off of the leading particles in the jets produced in $e^+e^-$ collisions.  Phenomenologically, this resummation would likely only be useful and necessary at future lepton colliders whose center-of-mass energies could far exceed the $Z$ pole.  Nevertheless, the smearing and softening of the cusp at fixed-order by all-order physics is clearly demonstrated, and we look forward to its application in a wide range of groomed jet analyses.

One obvious point of extension of our factorization theorem is to account for more than just one resolved emission that passes the jet groomer.  As mentioned earlier, the form of this cusp factorization theorem is similar to the resummation of non-global logarithms, in which there is no explicit measurement performed that establishes the number of resolved emissions.  Thus, one needs to sum over all mutually-exclusive number of resolved emissions and ensure that all degenerate limits are appropriately subtracted.  For example, the schematic form of the cross section contribution to the cusp region from two resolved soft emissions in one hemisphere would be
\begin{align}
&\frac{d\sigma^{\text{cusp, 2 emits}}}{d\rho}\\
&\hspace{0.5cm}=2H(Q^2)\sum_{1,2}\int d\vec k_1\, d\vec k_2\,d\rho_q\,d\rho_1\,d\rho_2\,d\rho_s\,\delta\left(\rho - 4k_1^+-4k_2^+-\rho_q-\rho_2-\rho_2-\rho_s \right)\,\Theta_\text{SD}\nonumber\\
&
\hspace{1cm}
\times\,H_s(\vec k_1,\vec k_2,\zcut)\,J_q(\rho_q)\, J_1(\rho_1)\, J_2(\rho_2)\,S_{q\bar q 12}(\rho_s)\int  d\tau_c\, d\tau_{sc}\, \Theta(\rho-\tau_c-\tau_{sc})\, J_{\bar q}(\tau_c)S_C(\tau_{sc})\,.\nonumber
\end{align}
Now, with two resolved emissions, 1 and 2, in the hemisphere, we need jet functions for both of them and to sum over their possible flavors.  The soft drop phase space constraints $\Theta_\text{SD}$ now are very complicated and all possible collinear limits of the two resolved emissions must be explicitly subtracted.  Additionally, the two resolved emissions could be in different hemispheres, which adds further complications.  Such a factorization theorem would first contribute at ${\cal O}(\alpha_s^2)$ order and so would be a subleading contribution to resummation at the cusp.  Nevertheless, to push the resummation of the cusp region to higher precision would require inclusion of contributions of this form.

A perhaps more relevant extension of our results here would be to jets produced at hadron colliders; specifically, jets defined through an algorithm with a fixed jet radius $R$.  Much of the structure of our factorization theorem would simply transfer, but, as with any analysis of jets in hadron collisions, the radiation outside of the jet is more challenging to control and contain.  Potentially event grooming along the lines of that proposed in \Ref{Baron:2020xoi} could help control out-of-jet radiation.  Additionally, jets at hadron colliders typically have a relatively small jet radius, like $R\sim 0.4$, and to good approximation that could be assumed to be parametrically smaller than 1.  Understanding the cusp region of jet groomers at hadron colliders may therefore require resummation of the observable, $\zcut$, and the jet radius $R$.  Further, because jets in hadron collisions can be either quark or gluon flavored, this introduces non-global contributions that affect the relative flavor fractions in the jet sample \cite{Frye:2016aiz}.  While more scales potentially means a more complicated factorization theorem or resummation structure, such an analysis could be directly compared to data that has already been collected and further improve our theoretical understanding of jets.

\acknowledgments

The authors thank Aditya Pathak for detailed comments about the manuscript and A.L.~thanks Simone Marzani for discussions about the transition region of groomers.  This work was facilitated in part by the Portland Institute for Computational Science and its resources acquired using NSF Grant DMS 1624776.

\appendix

\section{Anomalous Dimension Calculations}

To calculate these anomalous dimensions, we use dimensional regularization where $d=4-2\epsilon$ to render all phase space integrals finite and use modified minimal subtraction $\overline{\text{MS}}$ to eliminate terms with Euler-Mascheroni constants.

\subsection{Hard Function for Resolved Gluon Emission}\label{app:hardfunc}

The tree-level ``hard'' function for soft, wide-angle emission is
\begin{equation}
H_s^{(0)}(k^+,k^-) = \frac{\alpha_s}{\pi}C_F \frac{1}{k^+k^-}\,.
\end{equation}
At one-loop, its expression is \cite{Berends:1988zn,Catani:2000pi}
\begin{equation}
H_s^{(1)}(k^+,k^-) = -\frac{\alpha_s}{\pi}C_F \frac{1}{k^+k^-}
\frac{\alpha_s C_A}{4\pi}\left(
\log^2\frac{\mu^2}{k^+k^- Q^2}-\frac{5\pi^2}{6}
\right)\,.
\end{equation}
Because it starts at ${\cal O}(\alpha_s)$, its non-cusp anomalous dimension is proportional to the $\beta$-function:
\begin{equation}
\gamma_{H_s}^{(0)} = -\frac{\alpha_s}{2\pi}\beta_0\,.
\end{equation}

In addition to this virtual contribution to the anomalous dimension, with grooming it is also possible to that a real emission is groomed away.  In general, it might seem like we would have to use a correlated emission matrix element to describe the system in which one soft emission passes the groomer and one soft emission fails the groomer.  However, we restrict our accuracy to NLL, and the groomer forbids these two soft emissions from being collinear to one another because if they were, they would just be clustered together and pass the groomer.  So, the only divergences that are possible from which one-loop anomalous dimensions can be calculated are either when the emission that fails the groomer is parametrically lower energy than that which passes, and/or when the emission that fails is collinear to the anti-quark in the lighter hemisphere.  In either of these cases, the matrix element that describes such a configuration factorizes and is uncorrelated with the emission that passes.  For calculation beyond NLL, we would need the two-emission correlated matrix element, as that would affect the low-scale constants in this hard function.

With this understanding, the contribution to the hard function with grooming from the dipole $ij$ is
\begin{align}
H_{ij} &= -g^2\mu^{2\epsilon}{\bf T}_i\cdot {\bf T}_j\int\frac{d^dk}{(2\pi)^d}\frac{n_i\cdot n_j}{(n_i\cdot k )(n_j\cdot k)} \,2\pi \delta(k^2)\Theta(k^0)\\
&
\hspace{0.5cm}
\times\,\left[
\Theta(k^--k^+)
\Theta( k\cdot n_g-k^0n_g\cdot n_q )\Theta(k\cdot n_q-k^0n_g\cdot n_q )
\Theta\left(\zcut \frac{Q}{2}\left[2\frac{n\cdot k}{k^0}\right]^{\beta/2}-k^0\right)\right.\nonumber\\
&
\hspace{1cm}
\left.+\,
\Theta(k^+-k^-)\,\Theta\left(\zcut \frac{Q}{2}\left[2\frac{\bar n\cdot k}{k^0}\right]^{\beta/2}-k^0\right)
\right]\nonumber\\
&= -\frac{\alpha_s}{\pi}\mu^{2\epsilon}{\bf T}_i\cdot {\bf T}_j\frac{(4\pi)^\epsilon}{\pi^{1/2}\Gamma(1/2-\epsilon)}\int_0^\infty dk_\perp\, k_\perp^{-1-2\epsilon} \int_{-\infty}^\infty d\eta \int_0^{\pi} d\phi \,\sin^{-2\epsilon}\phi\,\frac{n_i\cdot n_j}{(n_i\cdot k )(n_j\cdot k)}\nonumber \\
&
\hspace{0.5cm}
\times\,\left[
\Theta(\eta)
\Theta\left(e^{-\eta}-\tan\theta\,\cos\phi\right)\Theta\left(\cot\frac{\theta}{2}-e^{\eta} \right)
\Theta\left(\zcut \frac{Q}{2}\left[2\frac{e^{-\eta}}{ \cosh\eta}\right]^{\beta/2}-k_\perp \cosh\eta\right)\right.\nonumber\\
&
\hspace{1cm}
\left.+\,
\Theta(-\eta)\,\Theta\left(\zcut \frac{Q}{2}\left[2\frac{e^{\eta}}{ \cosh\eta}\right]^{\beta/2}-k_\perp \cosh\eta\right)
\right]\,.\nonumber
\end{align}
Scaleless integrals have been explicitly removed.  Also, in the matrix element, the factors $(n_i\cdot k )(n_j\cdot k)$ have already had their dimensions through $k_\perp$ pulled out and made explicit in the factor $k_\perp^{-1-2\epsilon}$.  The phase space constraints here describe a parametrically soft gluon emitted off of the $q\bar q g$ final state that correspondingly fails the grooming algorithm and is not clustered with the resolved gluon that passes the groomer.

The integral over $k_\perp$ can therefore be done and one finds
\begin{align}
H_{ij} &= \frac{\alpha_s}{\pi}{\bf T}_i\cdot {\bf T}_j\frac{1}{\pi^{1/2}\Gamma(1/2-\epsilon)}\left(
\frac{\mu}{2^\beta\zcut Q}
\right)^{2\epsilon}\int_0^\infty d\eta\, \frac{e^{-2\epsilon\eta}}{2\epsilon}
(1+e^{2\eta})^{2\epsilon+\beta\epsilon} \int_0^{\pi} d\phi \,\sin^{-2\epsilon}\phi\nonumber \\
&
\hspace{2cm}
\times\,\frac{n_i\cdot n_j}{(n_i\cdot k )(n_j\cdot k)}\left[
\Theta\left(e^{-\eta}-\tan\theta\,\cos\phi\right)\Theta\left(\cot\frac{\theta}{2} -e^{\eta}\right)
+1\right]\,.
\end{align}
For the color-matrix dot products, recall that for $e^+e^-\to q\bar q g$ we have
\begin{equation}
{\bf T}_q\cdot {\bf T}_{\bar q} =\frac{C_A}{2}-C_F \,, \qquad
{\bf T}_q\cdot {\bf T}_g = -\frac{C_A}{2}\,, \qquad
{\bf T}_{\bar q}\cdot {\bf T}_g = -\frac{C_A}{2}\,.
\end{equation}

\subsubsection{$C_F$ Color Structure}

The $C_F$ color channel hard function is
\begin{align}
H_{C_F} &= -2\frac{\alpha_s}{\pi}C_F\frac{1}{\pi^{1/2}\Gamma(1/2-\epsilon)}\left(
\frac{\mu}{2^\beta\zcut Q}
\right)^{2\epsilon}\int_0^\infty d\eta\, \frac{e^{-2\epsilon\eta}}{2\epsilon}(1+e^{2\eta})^{2\epsilon+\beta\epsilon} \int_0^{\pi} d\phi \,\sin^{-2\epsilon}\phi\nonumber \\
&
\hspace{2cm}
\times\,\left[
\Theta\left(e^{-\eta}-\tan\theta\,\cos\phi\right)\Theta\left(\cot\frac{\theta}{2} -e^{\eta}\right)
+1\right]\,.
\end{align}
For the part of the integral with non-trivial phase space constraints, there is no collinear divergence.  Therefore, to evaluate the anomalous dimensions, we can everywhere set $\epsilon = 0$ in the integrand, only maintaining the $1/(2\epsilon)$ factor.  That is, we have the integral 
\begin{align}
A\left(
\theta
\right)&\equiv\int_0^\infty d\eta \int_0^{\pi} d\phi \,
\Theta\left(e^{-\eta}-\tan\theta\,\cos\phi\right)\Theta\left(\cot\frac{\theta}{2} -e^{\eta}\right)\\
&
=\left(
\pi - \cos^{-1}\left(
\frac{\tan\frac{\theta}{2}}{\tan\theta}
\right)
\right)\log \left(\cot\frac{\theta}{2}\right)-\cos^{-1}\left(
\frac{\tan\frac{\theta}{2}}{\tan\theta}
\right) \log\frac{\tan\theta}{2}
\nonumber\\
&
\hspace{1cm}
-\frac{1}{2}\text{Im}\, \text{Li}_2\left(
-e^{
-2 i \cos^{-1}\left(
\frac{\tan\frac{\theta}{2}}{\tan\theta}
\right)
}
\right)\nonumber\\
&
\hspace{1cm}
+\Theta(\tan\theta - 1)\left(
\frac{1}{2}\text{Im}\, \text{Li}_2\left(
-e^{
-2 i \cos^{-1}\left(
\cot\theta
\right)
}
\right)
+\cos^{-1}\left(
\cot\theta
\right) \log\frac{\tan\theta}{2}
\right)
\nonumber\,.
\end{align}
With the identification that
\begin{align}
&\tan\frac{\theta}{2} = \sqrt{\frac{k^+}{k^-}}\,, &\tan\theta = \frac{2\sqrt{k^+k^-}}{k^--k^+}\,,
\end{align}
this integral can be equivalently expressed as
\begin{align}
\label{eq:aint}
A\left(
\frac{k^+}{k^-}
\right)&=\frac{1}{2}\left(
\pi - \cos^{-1}\left(
\frac{k^--k^+}{2k^-}
\right)
\right)\log \left(\frac{k^-}{k^+}\right)-\cos^{-1}\left(
\frac{k^--k^+}{2k^-}
\right) \log\frac{\sqrt{k^+k^-}}{k^--k^+}
\nonumber\\
&
\hspace{1cm}
-\frac{1}{2}\text{Im}\, \text{Li}_2\left(
-e^{
-2 i \cos^{-1}\left(
\frac{k^--k^+}{2k^-}
\right)
}
\right)\nonumber\\
&
\hspace{1cm}
+\Theta\left(
2\sqrt{k^+k^-}-(k^--k^+)
\right)\nonumber\\
&\hspace{2cm}
\times\left(
\frac{1}{2}\text{Im}\, \text{Li}_2\left(
-e^{
-2 i \cos^{-1}\left(
\frac{k^--k^+}{2\sqrt{k^+k^-}}
\right)
}
\right)
+\cos^{-1}\left(
\frac{k^--k^+}{2\sqrt{k^+k^-}}
\right) \log\frac{\sqrt{k^+k^-}}{k^--k^+}
\right)
\nonumber\,.
\end{align}

The part of the hard function with trivial phase space constraints simply integrates to
\begin{align}
\int_0^\infty d\eta\, \frac{e^{-2\epsilon\eta}}{2\epsilon}(1+e^{2\eta})^{2\epsilon+\beta\epsilon} \int_0^{\pi} d\phi \,\sin^{-2\epsilon}\phi=-\frac{\pi}{4(1+\beta)\epsilon^2}-\frac{\pi}{\epsilon}\frac{\log\,2}{2(1+\beta)}\,.
\end{align}

Then, the contribution to the hard function's anomalous dimension from dropping soft, wide-angle particles in the $C_F$ channel is
\begin{align}
\gamma_{H_{C_F}}^{(0)} &=2\frac{\alpha_s C_F}{\pi}\left(
\frac{1}{1+\beta}\,\log\frac{\mu}{2^\beta \zcut Q}-\frac{A\left(
\frac{k^+}{k^-}
\right)}{\pi}
\right)\,.
\end{align}

\subsubsection{$C_A$ Color Structure}

The $C_A$ channel hard function is
\begin{align}
H_{C_A} &=-2\frac{\alpha_sC_A}{\pi}\frac{1}{\pi^{1/2}\Gamma(1/2-\epsilon)}\left(
\frac{\mu^2}{4^\beta\zcut^2 Q^2}
\right)^{\epsilon}\int_0^\infty d\eta\, \frac{e^{-2\epsilon\eta}}{2\epsilon}
(1+e^{2\eta})^{2\epsilon+\beta\epsilon} \int_0^{\pi} d\phi \,\sin^{-2\epsilon}\phi \\
&
\hspace{1cm}
\times\,\left[
\frac{\cos\phi\,\Theta\left(e^{-\eta}-\tan\theta\,\cos\phi\right)\Theta\left(\cot\frac{\theta}{2} -e^{\eta}\right)}{e^\eta\tan\frac{\theta}{2}+e^{-\eta}\cot\frac{\theta}{2}-2\cos\phi}
+\frac{\cos\phi}{e^{-\eta}\tan\frac{\theta}{2}+e^{\eta}\cot\frac{\theta}{2}-2\cos\phi}\right]\,.\nonumber
\end{align}
We have written this expression so that the pseudorapidity lies in $0\leq \eta<\infty$.  The two terms in the square brackets correspond to the matrix element times the phase space constraints in the right and left hemispheres, respectively, which is why the sign of $\eta$ in those terms differs.  By color conservation and phase space constraints, the explicit integrals are completely finite, so we can set $\epsilon = 0$ in all of the regulator terms.  We then have
\begin{align}
H_{C_A} &=-\frac{\alpha_sC_A}{\pi^2}\frac{1}{\epsilon}\left(
\frac{\mu^2}{4^\beta\zcut^2 Q^2}
\right)^{\epsilon}\int_0^\infty d\eta \int_0^{\pi} d\phi  \\
&
\hspace{1cm}
\times\,\left[
\frac{\cos\phi\,\Theta\left(e^{-\eta}-\tan\theta\,\cos\phi\right)\Theta\left(\cot\frac{\theta}{2} -e^{\eta}\right)}{e^\eta\tan\frac{\theta}{2}+e^{-\eta}\cot\frac{\theta}{2}-2\cos\phi}
+\frac{\cos\phi}{e^{-\eta}\tan\frac{\theta}{2}+e^{\eta}\cot\frac{\theta}{2}-2\cos\phi}\right]\,.\nonumber
\end{align}
Focusing first on the second term, with no phase space restrictions, we can just do this integral and find
\begin{align}
\int_0^\infty d\eta \int_0^{\pi} d\phi \,\frac{\cos\phi}{e^{-\eta}\tan\frac{\theta}{2}+e^{\eta}\cot\frac{\theta}{2}-2\cos\phi} = \frac{\pi}{2}\log\frac{\cos^2\frac{\theta}{2}}{\cos\theta}\,.
\end{align}

The first integral is much more challenging.  With the change of variables $z = e^{-\eta}$, it can be written over a compact domain as
\begin{align}
I(\theta)&\equiv \int_0^1 dz \int_0^\pi d\phi \, \frac{\cos\phi}{\tan\frac{\theta}{2}+z^2\cot\frac{\theta}{2}-2z\cos\phi}\, \Theta\left(
z - \tan \theta \cos\phi
\right)\Theta\left(
z-\tan\frac{\theta}{2}\right)
\,.
\end{align}
Note that
\begin{align}
&\tan\frac{\theta}{2} = \sqrt{\frac{k^+}{k^-}}\,, &\tan\theta = \frac{2\sqrt{k^+k^-}}{k^--k^+}\,.
\end{align}
This integral can also be expressed as
\begin{align}
I\left(\frac{k^+}{k^-}\right)&\equiv \int_0^1 dz \int_0^\pi d\phi \, \frac{\cos\phi}{\sqrt{\frac{k^+}{k^-}}+z^2\sqrt{\frac{k^-}{k^+}}-2z\cos\phi}\, \Theta\left(
z -\frac{2\sqrt{k^+k^-}}{k^--k^+} \cos\phi
\right)\Theta\left(
z-\sqrt{\frac{k^+}{k^-}}\right)
\,.
\end{align}
To very good approximation, we find that this integral can be expressed as:
\begin{align}
\label{eq:iint}
I\left(\frac{k^+}{k^-}\right)&\simeq 0.160-1.544\frac{k^+}{k^-}+0.332\sqrt{1.142-10.873\frac{k^+}{k^-}+27.192\left(\frac{k^+}{k^-}\right)^2}\,,
\end{align}
which is the equation for a hyperbola.  We will use this approximation in the evaluation of the cross section.

The contribution to the hard function's anomalous dimension from this color channel is thus
\begin{align}
\gamma_{H_{C_A}}^{(0)} &=-2\frac{\alpha_sC_A}{\pi}\left(
\frac{I\left(\frac{k^+}{k^-}\right)}{\pi}+\frac{1}{2}\log\frac{\cos^2\frac{\theta}{2}}{\cos\theta}
\right)\,.
\end{align}

\subsection{Global Soft Function}\label{app:softfunc}

Once the resolved gluon passes the groomer, all emissions that lie at smaller angle between that gluon and the hard (anti-)quark in the hemisphere pass the groomer with no constraints.  These emissions are described by a soft function and can be emitted off of any of the dipoles in the final state.  These emissions push the heavy hemisphere mass larger than just the contribution from the resolved gluon and to isolate their contribution, we consider the change in the mass:
\begin{equation}
\Delta m^2 = 2p_q\cdot (p_g'+p_s-p_g)\,.
\end{equation}
Here, $p_g$ is the gluon momentum without the additional soft emission and $p_g'$ is in the presence of the additional soft emission.  Because the soft drop algorithm employs Cambridge/Aachen clustering, we define the four vector $p_g'$ in the following way.  If the angle between the soft emission and the quark is less than to the gluon, we cluster the soft particle with the quark, and so we set $p_g' = p_g$.  On the other hand, if the angle from the soft particle to the gluon is less than the angle to the quark, we need $p_g'$ to recoil appropriately in absorbing the soft emission.

To construct the observable in this second case, we set
\begin{equation}
p_g = (E_g,0,0,E_g)\,,
\end{equation}
so that the quark momentum in this frame is
\begin{equation}
p_q = E_q(1,-\sin\theta_g,0,-\cos\theta_g)\,.
\end{equation}
The soft particle momentum is on-shell, but otherwise arbitrary:
\begin{equation}
p_s = E_s(1,\sin\theta\cos\phi,\sin\theta\sin\phi,\cos\theta)\,.
\end{equation}
Then, for the clustered momentum $p_g'+p_s$ to point in the same direction as $p_g$, we must have that
\begin{equation}
p_g' = (E_g',-E_s\sin\theta\cos\phi,-E_s\sin\theta\sin\phi,E_g-E_s\cos\theta)\,.
\end{equation}
Demanding that $p_g'$ is on-shell then fixes its energy to be
\begin{equation}
E_g' = \sqrt{E_g^2-2E_g E_s\cos\theta+E_s^2} \simeq E_g - 2E_s\cos\theta\,,
\end{equation}
where we have expanded to first-order in $E_s/E_g$.  Then, if the soft emission is closer to the gluon than the quark, the effect on the hemisphere mass is
\begin{equation}
\Delta m^2 \to 2p_q\cdot (p_g'+p_s-p_g) = 2E_q E_s (1-\cos\theta_{sg})\,.
\end{equation}

That is, the effect of a soft emission on the hemisphere jet mass with Cambridge/Aachen clustering is
\begin{align}
\rho_s = \frac{\Delta m^2}{E_q^2} = \Theta(\theta_{sg}-\theta_{sq})\,\frac{4E_s(1-\cos\theta_{sq})}{Q}+\Theta(\theta_{sq}-\theta_{sg})\,\frac{4E_s(1-\cos\theta_{sg})}{Q} \,.
\end{align}
Here, $\theta_{sq}$ and $\theta_{sg}$ are the angles between the soft emission and the quark or resolved gluon in the hemisphere, respectively.

\subsubsection{Calculation}

The contribution to the soft function $S_{q\bar q g}(\rho_s,\tau_s)$ from the dipole $ij$ is
\begin{align}
S_{ij} &= -g^2\mu^{2\epsilon}{\bf T}_i\cdot {\bf T}_j\int\frac{d^dk}{(2\pi)^d}\frac{n_i\cdot n_j}{(n_i\cdot k )(n_j\cdot k)} \,2\pi \delta(k^2)\Theta(k^0)\Theta(k^--k^+)\\
&
\hspace{1cm}
\times\,\left[
1-\Theta( k\cdot n_g-k^0n_g\cdot n_q )\Theta(k^+-k^0n_g\cdot n_q )
\right]\left[\Theta(n_g\cdot k-k^+)\,\delta\left(
\rho_s - \frac{4k^+}{Q}\right)\right.\nonumber\\
&
\hspace{2cm}
\left.+\Theta(k^+-k\cdot n_g)\,\delta\left(\rho_s-\frac{4k\cdot n_g}{Q}
\right)\right]\nonumber\\
&= -\frac{\alpha_s}{\pi}\mu^{2\epsilon}{\bf T}_i\cdot {\bf T}_j\frac{(4\pi)^\epsilon}{\pi^{1/2}\Gamma(1/2-\epsilon)}\int_0^\infty dk_\perp\, k_\perp^{-1-2\epsilon} \int_0^\infty d\eta \int_0^{2\pi} d\phi \,\sin^{-2\epsilon}\phi\,\frac{n_i\cdot n_j}{(n_i\cdot k )(n_j\cdot k)}\nonumber \\
&
\hspace{1cm}
\times\,
\left[
1-\Theta\left(e^{-\eta}-\tan\theta\,\cos\phi\right)\Theta\left(\cot\frac{\theta}{2}-e^{\eta} \right)
\right]\nonumber\\
&
\hspace{1cm}
\times\,\left[\Theta\left(\sinh\eta-\cot\frac{\theta_g}{2}\cos\phi\right)\,\delta\left(
\rho_s - \frac{4k_\perp e^{-\eta}}{Q}\right)\right.\nonumber\\
&
\hspace{1cm}
\left.+\Theta\left(\cot\frac{\theta_g}{2}\cos\phi-\sinh\eta\right)\,\delta\left(\rho_s-\frac{4k_\perp e^{-\eta}}{Q}\left(e^{2\eta}\sin^2\frac{\theta_g}{2}+\cos^2\frac{\theta_g}{2}-e^{\eta}\sin\theta_g \cos\phi\right)
\right)\right]\,.\nonumber
\end{align}
Scaleless integrals have been explicitly removed.  In particular, because soft emissions that are clustered in the groomed jet have no energy requirement on them, all integrals for emissions that land outside the clustered region are scaleless.  Also, in the matrix element, the factors $(n_i\cdot k )(n_j\cdot k)$ have already had their dimensions through $k_\perp$ pulled out and made explicit in the factor $k_\perp^{-1-2\epsilon}$.  The integral over $k_\perp$ can therefore be done and one finds
\begin{align}
S_{ij} &= -\frac{\alpha_s}{\pi}{\bf T}_i\cdot {\bf T}_j\frac{1}{\pi^{1/2}\Gamma(1/2-\epsilon)}\left(
\frac{4\mu}{Q}
\right)^{2\epsilon}\rho_s^{-1-2\epsilon}\int_0^\infty d\eta \,e^{-2\epsilon\eta}\int_0^{\pi} d\phi \,\sin^{-2\epsilon}\phi\,\frac{n_i\cdot n_j}{(n_i\cdot k )(n_j\cdot k)} \nonumber\\
&
\hspace{0cm}
\times\,
\Theta(\eta)\left[
\Theta\left(\tan\theta\,\cos\phi-e^{-\eta}\right)+\Theta\left(e^{-\eta}-\tan\theta\,\cos\phi\right)\Theta\left(e^{\eta}-\cot\frac{\theta}{2} \right)
\right]\\
&\hspace{1cm}
\times\left[\Theta\left(\sinh\eta-\cot\frac{\theta_g}{2}\cos\phi\right)\right.\nonumber\\
&
\hspace{2cm}
\left.+\Theta\left(\cot\frac{\theta_g}{2}\cos\phi-\sinh\eta\right)\,\left(e^{2\eta}\sin^2\frac{\theta_g}{2}+\cos^2\frac{\theta_g}{2}-e^{\eta}\sin\theta_g \cos\phi\right)^{2\epsilon}\right]\,.\nonumber
\end{align}

\subsubsection{$C_F$ Color Channel}

The $C_F$ color channel soft function is
\begin{align}
S_{C_F} &= 2\frac{\alpha_s}{\pi}C_F\frac{1}{\pi^{1/2}\Gamma(1/2-\epsilon)}\left(
\frac{4\mu}{Q}
\right)^{2\epsilon}\,\rho_s^{-1-2\epsilon}\int_0^\infty d\eta \, e^{-2\epsilon\eta} \int_0^{\pi} d\phi \,\sin^{-2\epsilon}\phi \\
&
\hspace{1cm}
\times\,\left(
1-\Theta\left(e^{-\eta}-\tan\theta\,\cos\phi\right)\Theta\left(\cot\frac{\theta}{2}-e^{\eta} \right)
\right)\nonumber\\
&\hspace{1cm}
\times\left[\Theta\left(\sinh\eta-\cot\frac{\theta_g}{2}\cos\phi\right)\right.\nonumber\\
&
\hspace{2cm}
\left.+\Theta\left(\cot\frac{\theta_g}{2}\cos\phi-\sinh\eta\right)\,\left(e^{2\eta}\sin^2\frac{\theta_g}{2}+\cos^2\frac{\theta_g}{2}-e^{\eta}\sin\theta_g \cos\phi\right)^{2\epsilon}\right]\,.\nonumber
\end{align}
Naively, this expression for the soft function is extremely complicated, due to the non-trivial phase space constraints from reclustering.  However, in the $C_F$ color channel, there is only a collinear singularity in the matrix element when $\eta \to \infty$, so for calculation of the anomalous dimension, we can set $\epsilon =0$ for the term in the brackets.  We will do this for which the soft function reduces to 
\begin{align}
S_{C_F} &= 2\frac{\alpha_s}{\pi}C_F\frac{1}{\pi^{1/2}\Gamma(1/2-\epsilon)}\left(
\frac{4\mu}{Q}
\right)^{2\epsilon}\,\rho_s^{-1-2\epsilon}\int_0^\infty d\eta \, e^{-2\epsilon\eta} \int_0^{\pi} d\phi \,\sin^{-2\epsilon}\phi \\
&
\hspace{1cm}
\times\,\left(
1-\Theta\left(e^{-\eta}-\tan\theta\,\cos\phi\right)\Theta\left(\cot\frac{\theta}{2}-e^{\eta} \right)
\right)\,.\nonumber
\end{align}

The non-trivial angular phase space constraints are exactly the same as that for the hard function $H_s$.  So, in that term, we can set $\epsilon =0$ and use the earlier result that
\begin{align}
&\int_0^\infty d\eta \int_0^{\pi} d\phi \,
\Theta\left(e^{-\eta}-\tan\theta\,\cos\phi\right)\Theta\left(\cot\frac{\theta}{2} -e^{\eta}\right)\\
&
\hspace{1cm}
=\left(
\pi - \cos^{-1}\left(
\frac{\tan\frac{\theta}{2}}{\tan\theta}
\right)
\right)\log \left(\cot\frac{\theta}{2}\right)-\cos^{-1}\left(
\frac{\tan\frac{\theta}{2}}{\tan\theta}
\right) \log\frac{\tan\theta}{2}
\nonumber\\
&
\hspace{2cm}
-\frac{1}{2}\text{Im}\, \text{Li}_2\left(
-e^{
-2 i \cos^{-1}\left(
\frac{\tan\frac{\theta}{2}}{\tan\theta}
\right)
}
\right)\nonumber\\
&
\hspace{2cm}
+\Theta(\tan\theta - 1)\left(
\frac{1}{2}\text{Im}\, \text{Li}_2\left(
-e^{
-2 i \cos^{-1}\left(
\cot\theta
\right)
}
\right)
+\cos^{-1}\left(
\cot\theta
\right) \log\frac{\tan\theta}{2}
\right)
\nonumber\,.
\end{align}
The term with no phase space constraints can easily be integrated and expanded in $\epsilon$:
\begin{equation}
\int_0^\infty d\eta \, e^{-2\epsilon\eta} \int_0^{\pi} d\phi \,\sin^{-2\epsilon}\phi =\frac{\pi}{2\epsilon}+\pi \log \,2\,.
\end{equation}

Then, the $C_F$ channel contribution to the Laplace-space soft function's anomalous dimensions are
\begin{align}
\gamma_{S_{C_F}}^{(0)} &=2\frac{\alpha_s C_F}{\pi}\left(
-\log\frac{4\tilde \rho \mu}{Q}+\frac{A\left(
\frac{k^+}{k^-}
\right)}{\pi}
\right)\,.
\end{align}

\subsubsection{$C_A$ Color Channel}

In the $C_A$ color channel, the emission matrix elements can be summed together to produce the soft function
\begin{align}
S_{C_A} &= 2\frac{\alpha_s}{\pi}C_A\frac{1}{\pi^{1/2}\Gamma(1/2-\epsilon)}\left(
\frac{16\mu^2}{Q^2}
\right)^{\epsilon}\,\rho_s^{-1-2\epsilon}\int_0^\infty d\eta \, e^{-2\epsilon\eta} \int_0^{\pi} d\phi \,\sin^{-2\epsilon}\phi \\
&
\hspace{1cm}
\times \frac{\cos\phi}{e^{\eta}\tan\frac{\theta}{2}+e^{-\eta}\cot\frac{\theta}{2}-2\cos\phi}\,\left(
1-\Theta\left(e^{-\eta}-\tan\theta\,\cos\phi\right)\Theta\left(\cot\frac{\theta}{2} -e^{\eta}\right)
\right)\nonumber\\
&\hspace{1cm}
\times\left[\Theta\left(\sinh\eta-2\cot\frac{\theta}{2}\cos\phi\right)\right.\nonumber\\
&
\hspace{2cm}
\left.+\Theta\left(2\cot\frac{\theta}{2}\cos\phi-\sinh\eta\right)\,\left(e^{2\eta}\sin^2\frac{\theta}{2}+\cos^2\frac{\theta}{2}-e^{\eta}\sin\theta \cos\phi\right)^{2\epsilon}\right]\,.\nonumber
\end{align}
Just as we did in the $C_F$ channel, this expression can be simplified for the purpose of calculating anomalous dimensions.  In the $C_A$ channel, there is only a collinear singularity when the soft emission becomes close to the gluon, so we can pull out the factor that is raised to the $\epsilon$ power that regulates the gluon collinear divergence in brackets, and then set $\epsilon = 0$ in the brackets.  The soft function then simplifies to
\begin{align}
S_{C_A} &= 2\frac{\alpha_s}{\pi}C_A\frac{2^{-2\epsilon}}{\pi^{1/2}\Gamma(1/2-\epsilon)}\left(
\frac{16\mu^2}{Q^2}
\right)^{\epsilon}\,\rho_s^{-1-2\epsilon}\int_0^\infty d\eta  \int_0^{\pi} d\phi \,\sin^{-2\epsilon}\phi\, \sin^{2\epsilon}\theta_g\\
&
\hspace{1cm}
\times \left(
1-\Theta\left(e^{-\eta}-\tan\theta\,\cos\phi\right)\Theta\left(\cot\frac{\theta}{2} -e^{\eta}\right)
\right)\nonumber\\
&\hspace{1cm}
\times\cos\phi\left(e^{\eta}\tan\frac{\theta_g}{2}+e^{-\eta}\cot\frac{\theta_g}{2}-2\cos\phi\right)^{-1+2\epsilon} \,.\nonumber
\end{align}

The integral with non-trivial phase space constraints is completely finite, and so we can set $\epsilon = 0$ in that term for calculating the anomalous dimensions.  The integral that remains is simply the $I(\frac{k^+}{k^-})$ integral identified in the calculation of the hard function.  For the integral with no phase space constraints, we change variables to
\begin{equation}
x = \cos\phi
\end{equation}
so the integral becomes
\begin{align}
&\sin^{2\epsilon}\theta\,\int_0^\infty d\eta  \int_{-1}^1 dx\,\left(1-x^2\right)^{-\frac{1}{2}-\epsilon}\,x\,\left(e^{\eta}\tan\frac{\theta}{2}+e^{-\eta}\cot\frac{\theta}{2}-2x\right)^{-1+2\epsilon} \\
&
\hspace{1cm}
=
\sin^{2\epsilon}\theta\,\frac{2\sqrt{\pi}\Gamma\left(\frac{3}{2}-\epsilon\right)}{\Gamma(2-\epsilon)}\int_{\tan\frac{\theta}{2}}^\infty dy\, \left|1-y\right|^{-1+2\epsilon}\left(
1+y
\right)^{-1+2\epsilon}\,y^{1-2\epsilon}\, \left(1+y^2\right)^{-1}\,\nonumber\\
&
\hspace{8cm}
\times \,_2F_1\left(
\frac{1}{2},1,2-\epsilon,\frac{4y^2}{\left(1+y^2\right)^2}
\right)
\,.\nonumber
\end{align}

Now, we can separate this integral into two regions: $y\in[\tan\frac{\theta}{2},1]$ and $y\in [1,\infty)$.  First, considering the interval $y\in [1,\infty)$, we make the change of variables
\begin{equation}
z = 1-\frac{1}{y}\,,
\end{equation}
so that we have the integral
\begin{align}
&\sin^{2\epsilon}\theta\,\frac{2\sqrt{\pi}\Gamma\left(\frac{3}{2}-\epsilon\right)}{\Gamma(2-\epsilon)}\int_0^1 dz\, z^{-1+2\epsilon}\,(1-z)^{1-2\epsilon}\,(2-z)^{-1+2\epsilon} \left(
1+(1-z)^2
\right)^{-1}\,\\
&
\hspace{8cm}
\times \,_2F_1\left(
\frac{1}{2},1,2-\epsilon,\frac{4(1-z)^2}{\left(1+(1-z)^2\right)^2}
\right)
\,.\nonumber
\end{align}
Now, we use the $+$-function expansion for the first factor in the integrand:
\begin{equation}
z^{-1+2\epsilon} = \frac{1}{2\epsilon}\delta(z)+\left(
\frac{1}{z}
\right)_++\cdots\,.
\end{equation}
We then find that this integral evaluates to
\begin{align}
[1,\infty) = \frac{\pi}{4\epsilon}+\frac{\pi}{2}\log(2\sin\theta)\,.
\end{align}

Now, we can tackle the other region.  We change variables to
\begin{equation}
z = \frac{1-y}{1-\tan\frac{\theta}{2}}\,,
\end{equation}
so that we have the integral
\begin{align}
&\frac{2\sqrt{\pi}\Gamma\left(\frac{3}{2}-\epsilon\right)}{\Gamma(2-\epsilon)}\left(1-\tan\frac{\theta}{2}\right)^{2\epsilon}\sin^{2\epsilon}\theta\int_0^1 dz \, z^{-1+2\epsilon}\,\left(2-\left(1-\tan\frac{\theta}{2}\right)z\right)^{-1+2\epsilon}  \\
&
\hspace{1cm}
\times \frac{\left(1-\left(1-\tan\frac{\theta}{2}\right)z\right)^{1-2\epsilon}}{1+\left(1-\left(1-\tan\frac{\theta}{2}\right)z\right)^2}\,_2F_1\left(
\frac{1}{2},1,2-\epsilon;\frac{4\left(1-\left(1-\tan\frac{\theta}{2}\right)z\right)^2}{\left(1+\left(1-\left(1-\tan\frac{\theta}{2}\right)z\right)^2\right)^2}
\right)\,.\nonumber
\end{align}
Doing the standard $+$-function expansion, we then find that this integral is
\begin{align}
\left[\tan\frac{\theta}{2},1\right] = \frac{\pi}{4\epsilon}+\frac{\pi}{2}\log(2\sin\theta)+\frac{\pi}{2}\log\left(1-\tan^2\frac{\theta}{2}\right)\,.
\end{align}
Putting it all together, we find that
\begin{align}
\int_0^\infty d\eta \, e^{-2\epsilon\eta} \int_0^{\pi} d\phi \,\sin^{-2\epsilon}\phi \frac{\cos\phi}{e^{\eta}\tan\frac{\theta}{2}+e^{-\eta}\cot\frac{\theta}{2}-2\cos\phi} = \frac{\pi}{2\epsilon}+\frac{\pi}{2}\log\left(8\cos\theta(1-\cos\theta)\right)\,.
\end{align}

With this result, the $C_A$ channel contribution to the soft function anomalous dimension is then
\begin{align}
\gamma_{S_{C_A}}&=2\frac{\alpha_s C_A}{\pi}\left(
-\log\frac{4 \tilde \rho\mu }{Q}-\frac{1}{2}\log\frac{(1-\cos\theta)\cos\theta}{2}+\frac{I\left(\frac{k^+}{k^-}\right)}{\pi}
\right)\,.
\end{align}

\subsection{Resolved Gluon Jet Function}\label{app:gluonjet}

To leading-order, the resolved gluon that passes the groomer is massless, and its contribution to the hemisphere mass comes exclusively from its energy and relative angle to the hard quark.  At higher-orders, the resolved gluon acquires a mass due to its own collinear splitting, and this will affect the hemisphere mass.  The contribution to the hemisphere jet mass from a collinear splitting of the soft gluon can be calculated as follows.  We first say that the gluon splits into two collinear particles, 1 and 2.  In the exactly collinear limit, these two particles' momenta just sum to the momentum of the gluon, $p_g$, so the new contribution to the hemisphere jet mass comes from their non-trivial splitting angle.  To leading power in the energy of the soft gluon, the new contribution to the hemisphere mass from the collinear splitting is
\begin{align}
\rho = \frac{\Delta m^2}{E_J^2} \equiv \frac{2 n\cdot(p_1+p_2-p_g)}{E_J}\,,
\end{align} 
where $p_g$ is the momentum of the gluon, $1$ and $2$ are its splitting products, and $n$ is the light-like direction of the hard quark.  In the collinear limit, we assume that the daughter particles 1 and 2 conserve the gluons' three-momentum; that is, collinear splittings do not affect the jet direction.  Further, the splitting products have a uniform azimuthal distribution about the gluon direction, so we can put their momentum in whatever direction is convenient.  We therefore express the momenta as
\begin{align}
p_g &= (E_g,0,0,E_g)\,,\\
p_1 &= (E_1,p_x,0,zE_g)\,,\\
p_2 &= (E_2,-p_x,0,(1-z)E_g)\,.
\end{align}
Here, $z$ is the fraction of the gluon's $z$-momentum that particle 1 takes away.  In this frame, note that the vector $n$ is
\begin{equation}
n = (1,-\sin\theta_g,0,-\cos\theta_g)\,.
\end{equation}
To lowest order in the splitting angle $\theta$, transverse momentum $p_x$ is just
\begin{equation}
p_x = z(1-z)E_g \theta\,.
\end{equation}
So, to put particles 1 and 2 on-shell, we must have that
\begin{align}
E_1 &= \sqrt{z^2 E_g^2 + z^2(1-z)^2E_g^2 \theta^2} \simeq zE_g + \frac{z(1-z)^2\theta^2}{2}E_g\,,\\
E_2 &= \sqrt{(1-z)^2 E_g^2 + z^2(1-z)^2E_g^2 \theta^2} \simeq (1-z)E_g + \frac{z^2(1-z)\theta^2}{2}E_g\,.
\end{align}
Then, the difference of energies before and after the splitting is
\begin{equation}
E_1+E_2-E_g \simeq \frac{z(1-z)\theta^2}{2}E_g\,.
\end{equation}
It then follows that the contribution to the hemisphere mass is
\begin{equation}
\rho = \frac{E_g}{E_J}z(1-z)\theta^2\,.
\end{equation}

The standard jet function mass expression is just $\rho = z(1-z)\theta^2$, so all one needs to do to calculate this jet function is make a change of argument to the logarithms that accounts for the small value of the gluon energy fraction.  In dimensionless light-cone coordinates, we express the energy of the gluon as
\begin{equation}
E_g = (k^-+k^+)E_J = \frac{k^-+k^+}{2}Q\,.
\end{equation}
Then, the new contribution to the hemisphere mass with light-cone coordinates is
\begin{equation}
\rho = (k^-+k^+)z(1-z)\theta^2
\end{equation}
With the rescaling of both the expression for the jet mass $\rho$ and of the jet energy $E_g$, the one-loop Laplace-space soft gluon jet function is then
\begin{equation}
\tilde J_g(\nu) = 1+ \frac{\alpha_s}{2\pi}  \left[ C_A \log^2\frac{4\mu^2 \nu}{(k^++k^-)Q^2} + \frac{\beta_0}{2} \, \log\frac{4\mu^2 \nu}{(k^++k^-)Q^2}+C_A\left(
\frac{67}{18}-\frac{\pi^2}{3}
\right)-\frac{10}{9}n_f T_R
\right]\,.
\end{equation}

\section{Summary of Anomalous Dimensions}\label{app:anomdims}

Here we summarize the one-loop anomalous dimensions of the functions in the factorization theorem of the soft drop grooming cusp.  The anomalous dimensions are presented in Laplace space, with $\tilde\rho$ and $\tilde\tau$ the Laplace-conjugates of the heavy and light hemisphere masses, respectively.  Further, we restrict to soft drop with $\beta = 0$ or the modified mass drop tagger groomer for simplicity, but the general results for the one-loop functions with arbitrary $\beta$ are presented elsewhere in this paper.
\begin{align}
\gamma_H &= -2\frac{\alpha_s C_F}{\pi} \left(\log\frac{\mu^2}{Q^2} + \frac{3}{2}\right)\,,\\
\gamma_{J_L} &=2\frac{\alpha_s C_F}{\pi}\left(
\log\frac{4\tilde \tau \mu^2}{Q^2}+\frac{3}{4}
\right) \,,\\
\gamma_{S_C} &= -2\frac{\alpha_s C_F}{\pi}\,\log\frac{4\tilde \tau \mu^2}{\zcut Q^2}\,,\\
\gamma_{J_H} &= 2\frac{\alpha_s C_F}{\pi}\left(
\log\frac{4\tilde \rho \mu^2}{Q^2}+\frac{3}{4}
\right)\,,\\
\gamma_{J_g} &=2\frac{\alpha_s}{\pi}\left(
C_A \log\frac{4\tilde \rho \mu^2}{(k^++k^-)Q^2} + \frac{\beta_0}{4}
\right) \,, \\
\gamma_{H_s} &= -2\frac{\alpha_s}{\pi}\left(
\frac{C_A}{2}\log\frac{\mu^2}{k^+k^- Q^2}+\frac{\beta_0}{4}-C_F\log\frac{\mu}{\zcut Q}+C_F \frac{A\left(\frac{k^+}{k^-}\right)}{\pi}+C_A \frac{I\left(\frac{k^+}{k^-}\right)}{\pi}\right.\\
&
\hspace{3cm}
\left.
+\frac{C_A}{2}\log\frac{k^-}{k^--k^+}
\right)\,,\nonumber \\
\gamma_{S} &= 2\frac{\alpha_s}{\pi}\left(
-(C_F+C_A)\log\frac{4\tilde \rho \mu}{Q}-\frac{C_A}{2}\log\frac{k^+(k^--k^+)}{(k^++k^-)^2}+C_F \frac{A\left(
\frac{k^+}{k^-}
\right)}{\pi}
+C_A \frac{I\left(
\frac{k^+}{k^-}
\right)}{\pi}\right)\,.
\end{align}
The integrals $A\left(\frac{k^+}{k^-}\right)$ and $I\left(\frac{k^+}{k^-}\right)$ are defined in \Eq{eq:aint} and \Eq{eq:iint}, respectively.  These anomalous dimensions sum to 0, demonstrating consistency of the factorization.

\subsection{Anomalous Dimensions with Boundary Factorization}\label{app:boundarysofts}

For generic momenta of the resolved gluon, where $k^+\sim k^-$, the anomalous dimensions presented above do indeed describe the cusp region of the groomed mass distribution.  However, this also includes the region in which $k^--k^+ \ll k^-$, when the resolved gluon approaches the hemisphere boundary.  In this limit, the collinear emissions off of the gluon may leak into the other hemisphere.  Measurement constraints are imposed on emissions in each hemisphere, either corresponding to  the heavy hemisphere mass or failing the grooming constraint, and if there is a hierarchy between these scales that can lead to large logarithms that must be resummed.  In particular, the resolved gluon's hard function $H_s$ and the soft function $S$ each have logarithms of the angle of the resolved gluon to the hemisphere boundary.  When $k^--k^+ \ll k^-$, these logarithms in the non-cusp anomalous dimensions themselves become large, and the functions must be refactorized to properly account for them.

This particular refactorization in this limit is exactly identical to the ``dressed gluon approximation'' presented in \Ref{Larkoski:2015zka}.  The anomalous dimensions for emissions in and out of the heavy hemisphere when the resolved gluon is near the boundary are, respectively,
\begin{align}
\gamma_\text{in} &= -\frac{\alpha_s C_A}{\pi}\log\frac{k^--k^+}{k^-}\,,\\
\gamma_\text{out} &= \frac{\alpha_s C_A}{\pi}\log\frac{k^--k^+}{k^-}\,.
\end{align}
Because these anomalous dimensions arise from the soft function $S$ or the resolved gluon's hard function $H_s$, their corresponding scales are set by these functions.  The new anomalous dimensions of $H_s$ and $S$ with this refactorization are
\begin{align}
\gamma_{H_s} &= -2\frac{\alpha_s}{\pi}\left(
\frac{C_A}{2}\log\frac{\mu^2}{k^+k^- Q^2}+\frac{\beta_0}{4}-C_F\log\frac{\mu}{\zcut Q}+C_F \frac{A\left(\frac{k^+}{k^-}\right)}{\pi}+C_A \frac{I\left(\frac{k^+}{k^-}\right)}{\pi}
\right)\,, \\
\gamma_{S} &= 2\frac{\alpha_s}{\pi}\left(
-(C_F+C_A)\log\frac{4\tilde \rho \mu}{Q}-\frac{C_A}{2}\log\frac{k^+k^-}{(k^++k^-)^2}+C_F \frac{A\left(
\frac{k^+}{k^-}
\right)}{\pi}
+C_A \frac{I\left(
\frac{k^+}{k^-}
\right)}{\pi}\right)\,.
\end{align}
These anomalous dimensions are then finite and well-defined for the entire phase space where $k^+\sim k^-$.

The in- and out-of-hemisphere anomalous dimensions can be used to solve the corresponding renormalization group evolution equations at NLL accuracy.  We have:
\begin{align}
S_\text{in}(\mu) &= \left(
1-\frac{k^+}{k^-}
\right)^{\frac{2C_A}{\beta_0}\log\frac{\alpha_s(\mu)}{\alpha_s(\mu_S)}}\,,\\
S_\text{out}(\mu) &=\left(
1-\frac{k^+}{k^-}
\right)^{-\frac{2C_A}{\beta_0}\log\frac{\alpha_s(\mu)}{\alpha_s(\mu_{H_s})}}\,,
\end{align}
where $\mu$ is the arbitrary renormalization scale and $\mu_S$ and $\mu_{H_s}$ are the scales for the appropriate functions.  Their product is independent of the renormalization scale $\mu$, and therefore is just a multiplicative factor to the cross section:
\begin{equation}
S_\text{in}(\mu)S_\text{out}(\mu)=\left(
1-\frac{k^+}{k^-}
\right)^{\frac{2C_A}{\beta_0}\log\frac{\alpha_s(\mu_{H_s})}{\alpha_s(\mu_S)}}\,.
\end{equation}
When included in the calculation of the cross section, this term will only contribute if it is raised to a positive exponent, corresponding to a suppression factor.  Thus, inclusion of this boundary term in the cross section takes the form:
\begin{align}\label{eq:boundarysuppfact}
\text{Boundary Factor}= \Theta(\mu_{H_s}-\mu_S)+\Theta(\mu_S-\mu_{H_s})\left(
1-\frac{k^+}{k^-}
\right)^{\frac{2C_A}{\beta_0}\log\frac{\alpha_s(\mu_{H_s})}{\alpha_s(\mu_S)}}\,.
\end{align}
That is, if $\mu_S>\mu_{H_s}$, a soft emission in the phase space region selected by the groomer would itself pass the groomer, which would be problematic for the hemisphere with no resolved emissions.  Such configurations should therefore be suppressed.  By contrast, if $\mu_S < \mu_{H_s}$ then a soft emission in the groomed phase space region would not have passed the groomer on its own, which then does not affect the hemisphere with no resolved emissions.  It is this factor in \Eq{eq:boundarysuppfact} that we include in the expression for NLL resummation of the cusp region of the groomer.

\section{Fixed-Coupling Expressions for Cusp Resummation}\label{app:fixed-coupling}

To account for the freezing of $\alpha_s$ below 1 GeV, we can expand the $K$ and $\omega$ functions in the limit that $\beta(\alpha_S) \to 0$. Freezing the coupling at the scale $\mu_0$, we first substitute the standard 1-loop expression for the running coupling,
\begin{equation}
    \alpha_s(\mu) = \frac{\alpha_s(\mu_0)}{1 + \frac{\alpha_s(\mu_0)}{2\pi}\beta_0 \log \frac{\mu}{\mu_0}},
\end{equation}
and then take $\beta_0 \to 0$ in the end. At NLL order, the $K$ function from \Eq{eq:NLL K function} becomes
\begin{equation}
    K(\mu, \mu_0) = \frac{\alpha_s}{4\pi}\left[ C_i \left( \Gamma_0 + \Gamma_1 \frac{\alpha_s}{4\pi} \right) \log^2 \frac{\mu}{\mu_0} + \gamma_0 \, \log \frac{\mu}{\mu_0} \right],
\end{equation}
while the $\omega$ function from \Eq{eq:NLL omega function} becomes
\begin{equation}
    \omega(\mu, \mu_0) = C_i \frac{\alpha_s}{4\pi}\left[ \Gamma_0 + \frac{\alpha_s}{4\pi}\Gamma_1 \right]\log \frac{\mu}{\mu_0}.
\end{equation}

\section{The Cusp for 1 TeV Collisions}\label{app:1tevcoll}

\begin{figure}
\begin{center}
\includegraphics[width=7cm]{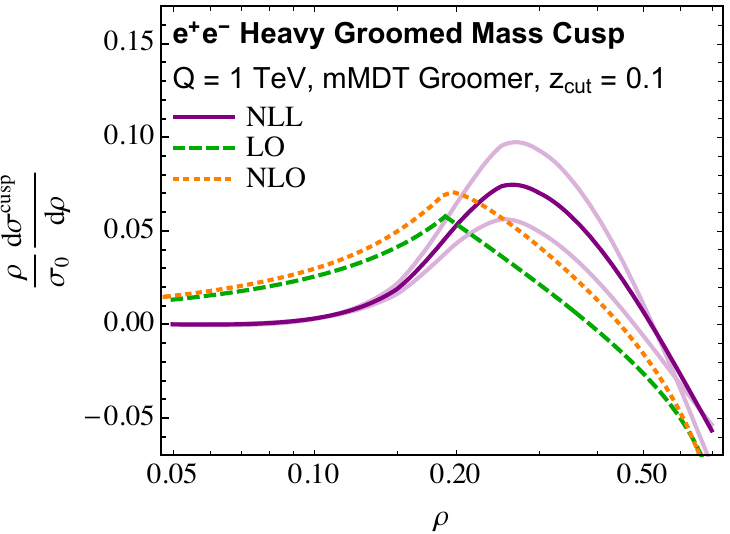}
\caption{
Comparison of the NLL cusp resummation to the cusp at leading- and next-to-leading fixed order in $e^+e^-$ collisions at 1 TeV.  The darker NLL distribution uses canonical scales, while the lighter curves below (above) vary the scale of the soft function up (down) by a factor of 2. 
\label{fig:lo_matched_1tev}
}
\end{center}
\end{figure}

In this appendix, we present plots of the NLL distribution of the cusp, \Eq{eq:resumnllfinal}, for $e^+e^-$ collisions at 1 TeV.  The differential cross section at the cusp is shown in \Fig{fig:lo_matched_1tev}, compared to the leading-order and next-to-leading order distribution about the cusp (i.e., with the limit $\rho\ll\zcut\ll1$ subtracted).  The lighter curves on the plot represent the variation of the scale of the soft function in the factorization theorem.  Now, at higher energies, scale variation mostly just affects the normalization of the cusp, and has little effect on its location.  Additionally, the location of the cusp in the NLL distribution is much closer to the cusp at fixed order than in the distributions at the $Z$ pole, but, importantly, is still shifted a bit to the right, toward values of $\rho$ above the location of the cusp at leading order, $\rho_\text{cusp} = 2\zcut$.

\bibliography{cusp}
\end{document}